\documentclass[twocolumn]{autart}    

\usepackage{cite}
\usepackage{amsmath,amssymb,amsfonts}
\usepackage{bm}
\usepackage{graphicx}
\usepackage{algorithm}
\usepackage{algpseudocode} 
\usepackage{hyperref}
\usepackage{textcomp}
\usepackage{xcolor}
\allowdisplaybreaks

\makeatletter
\g@addto@macro\normalsize{%
  \setlength\abovedisplayskip{0.5pt}
  \setlength\belowdisplayskip{0.5pt}
  \setlength\abovedisplayshortskip{0pt}
  \setlength\belowdisplayshortskip{0pt}
}
\makeatother

\begin{document}

\begin{frontmatter}

\title{Mitigating Overconfidence in Nonlinear Kalman Filters via Covariance Recalibration\thanksref{footnoteinfo}} 

\thanks[footnoteinfo]{The material in this paper was partially presented at the American Control Conference (ACC), July 8--12, 2024, Toronto, Canada. Corresponding author Shida Jiang Tel. +1 510-944-9339.}

\author[berkeley]{Shida Jiang}\ead{shida\_jiang@berkeley.edu},    
\author[berkeley]{Junzhe Shi}\ead{junzhe\_shi@berkeley.edu},               
\author[berkeley]{Scott Moura}\ead{smoura@berkeley.edu}  

\address[berkeley]{Department of Civil and Environmental Engineering, University of California, Berkeley, CA, USA}  


\begin{abstract}                          
The Kalman filter (KF) is an optimal linear state estimator for linear systems, and numerous extensions, including the extended Kalman filter (EKF), unscented Kalman filter (UKF), and cubature Kalman filter (CKF), have been developed for nonlinear systems. Although these nonlinear KFs differ in how they approximate nonlinear transformations, they all retain the same update framework as the linear KF. In this paper, we show that, under nonlinear measurements, this conventional framework inherently tends to underestimate the true posterior covariance, leading to overconfident covariance estimates. To the best of our knowledge, this is the first work to provide a mathematical proof of this systematic covariance underestimation in a general nonlinear KF framework. Motivated by this analysis, we propose a covariance-recalibrated framework that re-approximates the measurement model after the state update to better capture the actual effect of the Kalman gain on the posterior covariance; when recalibration indicates that an update is harmful, the update can be withdrawn. The proposed framework can be combined with essentially any existing nonlinear KF, and simulations across four nonlinear KFs and five applications show that it substantially improves both state and covariance estimation accuracy, often reducing errors by several orders of magnitude. The code and supplementary material are available at \url{https://github.com/Shida-Jiang/A-new-framework-for-nonlinear-Kalman-filters}.

\end{abstract}

\end{frontmatter}
\section{Introduction}
In control theory, the Kalman filter (KF) is a state estimation algorithm designed for linear systems. In such a system, a set of linear state transition functions describes how the states evolve, and a set of linear measurement functions describes the measured values of certain linear combinations of states. It is assumed that both the state transition and measurement functions contain some additive noise. Among all linear estimators, the KF optimally combines the system knowledge and measurements to minimize the expected value of the mean-squared error of the state estimation. Additionally, when the process noise, measurement noise, and state estimation errors all follow a multivariate normal (Gaussian) distribution, the state estimation given by the Kalman filter is also the maximum a posteriori estimate \cite{book3}. Due to these desirable features, KF is often referred to as the optimal linear state estimator. Since its birth in about 1960, KF has achieved massive success in navigation \cite{use1}, motion control \cite{use2}, statistics \cite{use3}, signal processing \cite{use4}, etc.

The most significant limitation of KF is that it can only be applied to systems with linear state transition functions and linear measurement functions. Unfortunately, many real-life systems are nonlinear \cite{book1}. Therefore, some extensions of KF have been developed to solve the state estimation problem for these nonlinear systems, and the most well-known ones are the extended Kalman filter (EKF) and the unscented Kalman filter (UKF). 

EKF linearizes the state-transition and measurement functions locally at each time step \cite{book1}. This transforms the nonlinear system into a locally linear one, to which the KF update can be applied. Although EKF is relatively simple, its first-order linearization neglects higher-order terms, which can lead to biased and overconfident state estimates. In particular, for a nonlinear function $f$, one generally has $\mathbb{E}[f(x)] \neq f(\mathbb{E}[x])$, and the resulting covariance estimate may substantially underestimate the true uncertainty. These issues are partially alleviated by the second-order extended Kalman filter (EKF2) \cite{book2}, which uses a second-order Taylor expansion. However, EKF2 requires Hessian information of the state-transition and measurement functions, which adds nontrivial computational complexity.

Classical EKF and EKF2 are usually formulated in a way that requires explicit Jacobian and, for EKF2, Hessian matrices, which may not always be available or convenient to compute in practice. For this reason, derivative-free alternatives have also been developed. In particular, the first- and second-order divided difference filters (DD1 and DD2) replace derivatives by divided differences and can be viewed as derivative-free counterparts of EKF and EKF2, respectively \cite{noderivative,norgaard2000new}. Closely related ideas also underlie the unscented Kalman filter (UKF), originally introduced by Julier and Uhlmann as a derivative-free extension of the KF to nonlinear systems \cite{UKF1,UKF2}. The UKF approximates the mean and covariance after a nonlinear transformation by propagating a set of deterministically chosen sigma points around the estimated state mean. These sigma points are selected so that their weighted mean and covariance match the prior mean and covariance, and the transformed moments are then approximated from the propagated points. As discussed in \cite{relation}, DD2 and the UKF are closely connected; under suitable conditions, they yield the same \emph{a priori} state estimate, although their covariance approximations may differ. Similar deterministic-sampling ideas have also been used in several other nonlinear KFs, including the cubature Kalman filter (CKF) \cite{CKF}, quadrature Kalman filter (QKF) \cite{QKF}, and square-root quadrature Kalman filter (SQKF) \cite{SQKF}. A disadvantage of these sampling-based nonlinear KFs is that the number of sampling points typically scales with the state dimension, which can make them increasingly more expensive than EKF for high-dimensional systems.

While different extensions of the Kalman filter (KF) approximate the first- and second-order moments of the state after a nonlinear transformation in different ways, they all retain the same algorithmic structure as the linear KF. In this paper, we refer to this structure as the ``conventional framework'' or ``old framework''. In each iteration, the estimation process consists of two stages: Predict and Update. In the ``predict'' step, the state mean and covariance are propagated through the process model using the estimates from the previous iteration. In the ``update'' step, the incoming measurement is used to correct the state prediction through a feedback term known as the Kalman gain. In the linear KF, this gain is analytically derived to minimize the trace of the a posteriori covariance matrix, and the corresponding covariance is then updated consistently with this optimal gain.

While this framework is the optimal linear state estimator for linear systems, applying it to nonlinear measurement functions often results in overconfident covariance estimates and degraded state accuracy. The fundamental issue is that the Kalman gain is computed as though it still minimizes the posterior covariance trace, even though this optimality no longer holds when the required moments are only approximated. As we will show later, the ``update'' step may even increase the estimation error in nonlinear systems. Consequently, the implicit assumption of optimality in the ``update'' step can be very harmful, because the filter becomes increasingly confident in estimates that are systematically biased or inaccurate.

The overconfidence phenomenon described above is well known in the literature. Graphical illustrations of EKF overconfidence have been provided in \cite{slam1}, and in the context of simultaneous localization and mapping (SLAM), researchers have repeatedly observed that nonlinear KFs tend to produce overconfident, rather than conservative, state estimates, especially after a long time \cite{slam2,slam3,slam4}. Moreover, this overconfidence issue becomes more pronounced when the measurement noise variance is small \cite{theoretical,consistency}, because the linearization (or other approximation) error then constitutes a larger fraction of the total error budget, making the filter’s covariance underestimate more severe. In some cases, this leads to a highly undesirable and counterintuitive behavior: the actual estimation error increases as the measurement noise decreases. The underlying reason is that the actual estimation errors decrease much more slowly than the estimated standard deviations, so the filter becomes overconfident and is less able to correct itself in later iterations. As sensor technology continues to improve and measurement noise is further reduced across many applications, we can reasonably expect this problem to become even more severe if we continue to rely on the conventional framework for nonlinear KFs.

Unfortunately, so far, there is no general solution that can effectively prevent nonlinear KFs from overconfidence while also improving state estimation accuracy. Existing approaches mainly reduce approximation error either by using higher-order local approximations (e.g., EKF2 instead of EKF) or by using stochastic integration rules that become asymptotically exact as the number of integration samples increases \cite{dunik2013stochastic}. However, the former only mitigates the problem, typically at substantially higher computational cost, while the latter still provides only approximate moment evaluations under a finite computational budget. As a result, the overconfidence problem generally remains. The lack of theoretical analysis for nonlinear KFs makes finding a general solution even harder. In the few existing theoretical analyses \cite{relation, theoretical, theoretical2}, the results are based on the assumption that third- and higher-order terms in the Taylor expansion of the measurement function can be ignored. This assumption may not be reasonable because the predicted states can be far from the updated states, and neglecting higher-order terms can lead to significant errors in the ``update'' step.

Apart from more accurate approximations, another widely used strategy for mitigating filter overconfidence is covariance inflation \cite{inflation1,inflation3}. 
In this approach, the estimated state covariance is deliberately enlarged, and the associated inflation factor is adjusted to compensate for under-dispersion and errors caused by nonlinearity. While covariance inflation can alleviate overconfidence, its performance depends heavily on the choice of the inflation factor, whose value typically requires extensive fine-tuning. This tuning can be computationally costly and problem-dependent, thereby undermining the universality of the approach \cite{inflation4,inflation5}.

In this study, we propose a covariance-recalibrated framework for nonlinear Kalman filters to address the problem of overconfidence. The key idea is to introduce an additional step that reassesses the actual effect of the Kalman gain after the state update. With this step, the state covariance can be estimated more accurately, and harmful updates can be withdrawn when necessary. The main contributions of this paper are as follows:
\begin{itemize}
    \item To the best of our knowledge, this is the first work to provide a mathematical proof that the conventional nonlinear KF framework, which has been used for about sixty years, generally tends to underestimate the trace of the state covariance matrix.
    \item We propose a new framework to address inaccurate covariance estimation, and we validate its effectiveness through both rigorous theoretical analysis and empirical results.
    \item The proposed framework can be combined with almost any type of nonlinear KF, reducing state estimation error by more than an order of magnitude across various applications when the measurement noise is low.
\end{itemize}

The rest of the paper is organized as follows. Section 2 conducts a rigorous theoretical analysis of the problem with the conventional framework for nonlinear KFs and introduces a new framework that solves it. Section 3 presents a comprehensive comparison between the two frameworks. Specifically, both frameworks are combined with four different types of nonlinear KFs and validated on five distinct applications. The conclusions are drawn in Section 4.
\section{Method}
\subsection{General formulation of nonlinear Kalman filters}
Consider a discrete-time system with nonlinear dynamics
\begin{equation}\label{setup}
    x_k=f(x_{k-1},u_{k-1})+w_{k-1},
\end{equation}
where $x_k$ denotes the (unknown) true states at time step $k$, $u_{k-1}$ is the known input applied between time steps $k-1$ and $k$, and $w_{k-1}$ is the process noise, assumed zero-mean with covariance $Q_{k-1}$, and independent of $x_{k-1}$ and past measurements. At time step $k-1$, the filter maintains an a posteriori state estimate $\hat{x}_{k-1|k-1}$ with associated error covariance $P_{k-1|k-1}$. 

For any nonlinear KF variant, the first step of each iteration is the ``predict'' step, which propagates $\hat{x}_{k-1|k-1}$ and $P_{k-1|k-1}$ through the process model \eqref{setup} to obtain the a priori quantities $\hat{x}_{k|k-1}$ and $P_{k|k-1}$. The predicted state estimate at time step $k$ is denoted by $\hat{x}_{k|k-1}$, and the corresponding estimation error $x_k-\hat{x}_{k|k-1}$ is assumed to follow a zero-mean multivariate distribution with covariance $P_{k|k-1}$. Formally,
\begin{equation}\label{conditional}
    \begin{cases}
        \mathbb{E}[x_k-\hat{x}_{k|k-1}|z_{1:k-1}]=0,\\
        P_{k|k-1}=\mathrm{Var}(x_k - \hat{x}_{k|k-1}|z_{1:k-1}),
    \end{cases}
\end{equation}
where $z_{1:k-1}$ denotes the past measurements. Here and throughout the paper, the subscript $k|k-1$ denotes the one-step-ahead predicted quantity at time $k$ conditioned on the observation history $z_{1:k-1}$, whereas $k|k$ denotes the updated quantity after observing $z_k$. In the subsequent analysis, we work conditionally on the observation history. In particular, for any fixed realization of $z_{1:k-1}$, 
$P_{k|k-1}$ is a deterministic function of that realization; likewise, $P_{k-1|k-1}$ is deterministic given $z_{1:k-1}$. Unless stated otherwise, we suppress the conditioning in the notation for all means, covariances, and cross-covariances (e.g., $P_{k|k-1}$, $P_{z,k}$, $P_{xz,k}$, and $S_k$, which will be formally defined later).

The second step of each iteration for a nonlinear KF is called the ``update'', which corrects the predicted state based on the new measurement. The measurement model is
\begin{equation}\label{zk}
    z_k=h(x_k)+v_k,
\end{equation}
where $z_k$ is the measurement at time step $k$, and $v_k$ is the measurement noise, assumed zero-mean with covariance $R_k$. Given the predicted state, the measurement can be predicted (using a specific nonlinear KF algorithm), and we denote this predicted measurement by $\hat{z}_{k|k-1}$. The corresponding (noise-free) prediction error $h(x_k)-\hat{z}_{k|k-1}$ is assumed zero-mean with covariance $P_{z,k}$. The cross-covariance between the state prediction error and the measurement prediction error is denoted by $P_{xz,k}$.

The difference between the actual and predicted measurements is called the measurement residual (or innovation), denoted by
\begin{equation}\label{resi}
    \tilde{z}_k:=z_k-\hat{z}_{k|k-1}.
\end{equation}
Since both $v_k$ and $h(x_k)-\hat{z}_{k|k-1}$ are zero-mean, the innovation $\tilde{z}_k$ is also zero-mean. If, in addition, the measurement noise $v_k$ is assumed to be independent of the (noise-free) prediction error $h(x_k)-\hat{z}_{k|k-1}$, the covariance of the innovation is
\begin{equation} \label{residual_cov}
    S_{k}=P_{z,k}+R_k.
\end{equation}
The KF update for the state estimate is
\begin{equation}\label{state_update}
    \hat{x}_{k|k}=\hat{x}_{k|k-1}+K_k\tilde{z}_k,
\end{equation}
where $\hat{x}_{k|k}$ is the updated state estimate and $K_k$ is the Kalman gain at time step $k$. Using \eqref{resi}--\eqref{state_update}, the covariance of the updated estimation error, conditional on the past measurements $z_{1:k-1}$, can be written as \cite{josephUKF}
\begin{equation}\label{state_variance_true}
    P_{k|k} = P_{k|k-1} + K_k S_k K_k^T - P_{xz,k} K_k^T - K_k P_{xz,k}^T.
\end{equation}
Equation~\eqref{state_variance_true} is a generalized Joseph form that applies to any nonlinear KF variant. For a linear or linearized measurement model with measurement matrix $C_k$, where $P_{xz,k}=P_{k|k-1}C_k^T$ and $S_k=C_kP_{k|k-1}C_k^T+R_k$, expanding the classical expression $(I-K_kC_k)P_{k|k-1}(I-K_kC_k)^T+K_kR_kK_k^T$ recovers \eqref{state_variance_true} exactly.
The Kalman gain $K_k$ is chosen to minimize $\mathrm{tr}(P_{k|k})$. By differentiating $\mathrm{tr}(P_{k|k})$ in \eqref{state_variance_true} with respect to $K_k$, we obtain the optimal gain
\begin{equation}\label{Kalmangain}
    K_{k,\text{op}}=P_{xz,k}S_k^{-1},
\end{equation}
where $K_{k,\mathrm{op}}$ denotes the optimal Kalman gain. Substituting \eqref{Kalmangain} into \eqref{state_variance_true} yields the well-known expression
\begin{equation}\label{simple_covariance}
    P_{k|k,\mathrm{op}} = P_{k|k-1} - K_{k,\mathrm{op}} S_k K_{k,\mathrm{op}}^T.
\end{equation}
Equations \eqref{state_update}–\eqref{simple_covariance} summarize the key principle of KF: the state estimate is updated in such a way that the trace of the a posteriori error covariance is minimized. However, for systems with nonlinear measurement functions $h(\cdot)$, these expressions involve covariance terms $P_{k|k-1}$, $P_{z,k}$, $S_k$, and $P_{xz,k}$ that, in general, are defined as expectations with respect to the full distribution of the state and cannot be uniquely determined from the first two moments alone. Since nonlinear KFs propagate only the mean and covariance of the state, this information is generally insufficient to recover the true values of these matrices, so they must be approximated. In fact, as we will show next, inaccurate approximations regarding these matrices are the key factors that can make the conventional framework for nonlinear KFs overconfident and inaccurate.

Before concluding this subsection, we note that the formulation
above extends directly to non-additive noise. Specifically, if the
process noise enters the dynamics as
$x_k = f(x_{k-1}, u_{k-1}, w_{k-1})$, we can define the prediction-time
augmented variable
$\chi_{k-1} := [x_{k-1}^T \; w_{k-1}^T]^T$. Likewise, if the
measurement noise is non-additive, $z_k = h(x_k, v_k)$, we can define
the update-time augmented variable
$\eta_k := [x_k^T \; v_k^T]^T$. Since the first two moments of
these augmented variables are determined by the first two moments of
$x_{k-1}$, $w_{k-1}$, and $v_k$, the same algorithmic machinery (e.g.,
linearization or sigma-point propagation) applies to $\chi_{k-1}$ and
$\eta_k$ to approximate the first two moments of $x_k$ and $z_k$.
Importantly, $w_{k-1}$ and $v_k$ are not treated as persistent states;
they are i.i.d.\ zero-mean draws at each step and are included in the
augmentation only to capture their nonlinear effect on the moments.

\subsection{The conventional framework and its problem}
While calculating the exact values of $P_{k|k-1}, P_{z,k}$, $S_k$, and $P_{xz,k}$ is generally impossible by just using the first two moments of the states, it is possible to make some approximations. Without loss of generality, this problem of approximation can be formulated as follows.

\textbf{The general approximation problem in nonlinear Kalman filters:}
Given an estimated mean $\hat{x} \approx \mathbb{E}[x]$ and an estimated covariance $\textcolor{blue}{P_{x,\mathrm{est}}} \approx \mathrm{Var}(x)$ of a random vector $x$, a nonlinear KF seeks to approximate the mean and covariance of the transformed variable $g(x)$, as well as the cross-covariance between $x$ and $g(x)$. In this section, covariance matrices obtained through nonlinear KF approximation are distinguished from the corresponding actual covariances by their subscripts (``est'' or an explicit conditioning index such as $k|k-1$); where color is available, they are additionally shown in blue. Their formal probabilistic interpretation will be introduced later. These approximations can be generally written as
\begin{equation}
\begin{cases}
\Phi_e\bigl(g;\hat{x},\textcolor{blue}{P_{x,\mathrm{est}}}\bigr) \approx \mathbb{E}[g(x)],\\[4pt]
\textcolor{blue}{P_{g,\mathrm{est}}} := \Phi_{\mathrm{var}}\bigl(g;\hat{x},\textcolor{blue}{P_{x,\mathrm{est}}}\bigr) \approx \mathrm{Var}(g(x)),\\[4pt]
\textcolor{blue}{P_{xg,\mathrm{est}}} := \Phi_{\mathrm{cov}}\bigl(g;\hat{x},\textcolor{blue}{P_{x,\mathrm{est}}}\bigr) \approx \mathrm{Cov}(x,g(x)),
\end{cases}
\end{equation}
where $\Phi_e$, $\Phi_{\mathrm{var}}$, and $\Phi_{\mathrm{cov}}$ denote the approximation mappings determined by the chosen nonlinear KF (e.g., EKF, UKF, CKF). Here, $g$ may represent either the state transition function $f$ in the prediction step or the measurement function $h$ in the update step, so this approximation problem is typically solved at least twice in each iteration of a nonlinear KF. Specifically, $P_{k|k-1}$, $P_{z,k}$, $S_k$, and $P_{xz,k}$ can be respectively approximated by
\begin{equation}\label{approx1}
\begin{cases}
\textcolor{blue}{P_{k|k-1,\mathrm{est}}} := \Phi_{\mathrm{var}}(f;\hat{x}_{k-1|k-1},\textcolor{blue}{P_{k-1|k-1,\mathrm{est}}}) + Q_{k-1},\\
\textcolor{blue}{P_{z,k|k-1}} := \Phi_{\mathrm{var}}(h;\hat{x}_{k|k-1},\textcolor{blue}{P_{k|k-1,\mathrm{est}}}),\\
\textcolor{blue}{S_{k|k-1}} := \textcolor{blue}{P_{z,k|k-1}} + R_k,\\
\textcolor{blue}{P_{xz,k|k-1}} := \Phi_{\mathrm{cov}}(h;\hat{x}_{k|k-1},\textcolor{blue}{P_{k|k-1,\mathrm{est}}}).
\end{cases}
\end{equation}
The subscript $k|k-1$ indicates that these covariance matrices are computed based on the predicted state estimate $\hat{x}_{k|k-1}$.

Although the corresponding $\Phi_{e},\Phi_{\mathrm{var}}$, and $\Phi_{\mathrm{cov}}$ differ, all existing types of nonlinear KFs directly replace $P_{k|k-1}$, $S_k$, and $P_{xz,k}$ in (\ref{Kalmangain}) and (\ref{simple_covariance}) with the approximated values in (\ref{approx1}). Namely, the Kalman gain is selected as
\begin{equation}\label{Kalmangain_real}
    \textcolor{blue}{K_{k,\text{est}}}:=\textcolor{blue}{P_{xz,k|k-1}}\textcolor{blue}{S_{k|k-1}^{-1}},
\end{equation}
and the state covariance matrix is updated by
\begin{equation}\label{oldkf}
\begin{aligned}
    \textcolor{blue}{P_{k|k,\text{est}}}&:=\textcolor{blue}{P_{k|k-1,\text{est}}}-\textcolor{blue}{K_{k,\text{est}}}\textcolor{blue}{S_{k|k-1}}\textcolor{blue}{K_{k,\text{est}}^T}\\
    &=\textcolor{blue}{P_{k|k-1,\text{est}}}-\textcolor{blue}{P_{xz,k|k-1}}\textcolor{blue}{S_{k|k-1}^{-1}}\textcolor{blue}{P_{xz,k|k-1}^T}.
\end{aligned}
\end{equation}
On the other hand, the actual value of the updated state covariance matrix $\textcolor{blue}{P_{k|k,\text{ac}}}$ given the selected Kalman gain in (\ref{Kalmangain_real}) can be calculated by substituting (\ref{Kalmangain_real}) into (\ref{state_variance_true}). Namely,
\begin{equation}\label{P_true}
\begin{aligned}
    &\textcolor{blue}{P_{k|k,\text{ac}}}=P_{k|k-1}+\textcolor{blue}{P_{xz,k|k-1}}\textcolor{blue}{S_{k|k-1}^{-1}}S_{k}\textcolor{blue}{S_{k|k-1}^{-1}}\textcolor{blue}{P_{xz,k|k-1}^T}\\
    &\quad -P_{xz,k}\textcolor{blue}{S_{k|k-1}^{-1}}\textcolor{blue}{P_{xz,k|k-1}^T}-\textcolor{blue}{P_{xz,k|k-1}}\textcolor{blue}{S_{k|k-1}^{-1}}P_{xz,k}^T.
\end{aligned}
\end{equation}
Note that $\textcolor{blue}{P_{k|k,\text{ac}}}$ plays the role of $P_{k|k}$ in (\ref{state_variance_true}), and $\textcolor{blue}{P_{k|k,\text{est}}}$ plays the role of $P_{k|k,\text{op}}$ in (\ref{simple_covariance}).
Intuitively, tr$(\textcolor{blue}{P_{k|k,\text{ac}}})$ tends to be larger than tr$(\textcolor{blue}{P_{k|k,\text{est}}})$ since tr$(P_{k|k,\text{op}}) = \min \ $tr$(P_{k|k})\leq $ tr$(P_{k|k})$. However, this relationship of size is not always preserved after the approximations. 
For example, substituting $\textcolor{blue}{P_{k|k-1,\text{est}}}=P_{k|k-1}$, $P_{xz,k}=S_k=1$, and $\textcolor{blue}{P_{xz,k|k-1}}=\textcolor{blue}{S_{k|k-1}}=0.9$ into (\ref{oldkf}) and (\ref{P_true}) yields $\textcolor{blue}{P_{k|k,\text{est}}}-\textcolor{blue}{P_{k|k,\text{ac}}}=0.1>0$. 
Nevertheless, in most cases in practice \cite{over1,over2}, we do find that tr$(\textcolor{blue}{P_{k|k, \text{ac}}})>\textrm{tr}(\textcolor{blue}{P_{k|k, \text{est}}})$, meaning that using (\ref{oldkf}) will often (although not always) underestimate the actual covariance. 

To formalize the preceding discussion, we interpret the approximated covariance matrices
$\textcolor{blue}{P_{z,k|k-1}}$, $\textcolor{blue}{S_{k|k-1}}$, and
$\textcolor{blue}{P_{xz,k|k-1}}$ produced by a nonlinear KF as
(random) estimators of the actual quantities $P_{z,k}$, $S_k$, and $P_{xz,k}$.
More precisely, we regard $\textcolor{blue}{P_{xz,k|k-1}}$ as a random matrix,
$\textcolor{blue}{P_{z,k|k-1}}$ as a random positive semidefinite matrix, and
$\textcolor{blue}{S_{k|k-1}}$ as a random positive definite matrix. As noted above, these approximated covariance matrices are distinguished from the corresponding actual covariances by their subscripts. Additionally, we assume that these approximated covariance matrices are unbiased estimators and satisfy
\begin{equation}\label{expcondition}
    \begin{cases} \mathbb{E}[\textcolor{blue}{P_{z,k|k-1}}] = P_{z,k}, \\ 
    \mathbb{E}[\textcolor{blue}{S_{k|k-1}}] =S_k, \\ 
    \mathbb{E}[\textcolor{blue}{P_{xz,k|k-1}}]= P_{xz,k}.\end{cases} 
\end{equation}

Here, randomness models uncertainty in the moment approximations, even for a deterministic moment rule, and the expectations are taken with the corresponding true covariance matrices held fixed.

These additional definitions and assumptions are introduced to facilitate consideration of the ``average'' case and provide a more mathematically rigorous definition of the overconfidence caused by (\ref{oldkf}). Specifically, (\ref{expcondition}) should be understood as follows. If a systematic, nonzero-mean bias in these approximations were known a priori, it would be natural to modify the filter so as to subtract this bias, and to analyze only the residual approximation error. Thus, it is reasonable to model the remaining covariance errors as zero-mean random matrices. Note that this viewpoint is conceptually analogous to the standard treatment of deterministic biases in augmented-state or separate-bias Kalman filters~\cite{bias,bias2}, where unknown biases are initialized as zero-mean random variables. Namely, any known mean component is absorbed into the system model, and only the zero-mean uncertainty is left for the filter to estimate.

With (\ref{expcondition}), we have the following theorem about the overconfidence of nonlinear Kalman filters.
\begin{thm}\label{theorem1}
Suppose that $\textcolor{blue}{P_{xz,k|k-1}}$ and $\textcolor{blue}{S_{k|k-1}}$ are random matrices that satisfy (\ref{expcondition}). $\textcolor{blue}{S_{k|k-1}}$ is symmetric and positive definite. Further assume that
\begin{equation}\label{E1}
    P_{k|k-1}\succeq \mathbb{E}[\textcolor{blue}{P_{k|k-1,\mathrm{est}}}]
\end{equation}
where ``$\succeq$'' denotes the positive semidefinite ordering.
Then,
\begin{equation}\label{theorem}
    \mathbb{E}[\textcolor{blue}{P_{k|k,\mathrm{ac}}} - \textcolor{blue}{P_{k|k,\mathrm{est}}}]\succeq 0,
\end{equation}
and the equality only holds when $\mathbb{E}[\textcolor{blue}{P_{k|k-1,\mathrm{est}}}]= P_{k|k-1}$ and $\textcolor{blue}{P_{xz,k|k-1}}\textcolor{blue}{S_{k|k-1}^{-1}}$ is constant almost surely. Note that this theorem does not require $\textcolor{blue}{P_{xz,k|k-1}}$ and $\textcolor{blue}{S_{k|k-1}}$ to be independent.
\end{thm}

\begin{pf}
Let $\textcolor{blue}{P}=\textcolor{blue}{P_{xz,k|k-1}}S_k^{-\frac{1}{2}}$, $\textcolor{blue}{S}=S_k^{-\frac{1}{2}}\textcolor{blue}{S_{k|k-1}}S_k^{-\frac{1}{2}}$. According to (\ref{expcondition}),
\begin{equation}\label{expcondition2}
    \begin{cases}
        \mathbb{E}[\textcolor{blue}{P}] = \bar{P}= P_{xz,k}S_k^{-\frac{1}{2}}, \\
        \mathbb{E}[\textcolor{blue}{S}] = I.
    \end{cases}
\end{equation}
Then, according to (\ref{oldkf}),
\begin{equation}\label{equ1}
\begin{aligned}
    &\textcolor{blue}{P_{k|k,\mathrm{ac}}} - \textcolor{blue}{P_{k|k,\mathrm{est}}} -(P_{k|k-1}-\textcolor{blue}{P_{k|k-1,\mathrm{est}}}) \\
    =&\,\textcolor{blue}{P}\textcolor{blue}{S^{-2}P^T}
    -\bar{P}\textcolor{blue}{S^{-1}P^T}
    -\textcolor{blue}{PS^{-1}}\bar{P}^T
    +\textcolor{blue}{PS^{-1}P^T} \\
    =&\,(\textcolor{blue}{PS^{-1}}-\bar{P})(\textcolor{blue}{PS^{-1}}-\bar{P})^T
    +\textcolor{blue}{PS^{-1}P^T}-\bar{P}\bar{P}^T.
\end{aligned}
\end{equation}
For any fixed unit vector $\bm{v}$, $\textcolor{blue}{P^T}\bm{v}$ is a random vector with a mean of $\bar{P}^T\bm{v}$. According to Lemma \ref{lemmakey} in Appendix \ref{appendix_proof},
\begin{equation}
    \mathbb{E}[\bm{v}^T\textcolor{blue}{P}\textcolor{blue}{S^{-1}}\textcolor{blue}{P^T}\bm{v}]
    \geq \bm{v}^T\bar{P}\bar{P}^T\bm{v}.
\end{equation}
Moving the right-hand side to the left, we get
\begin{equation}\label{eqn:lemma5-psd}
    \mathbb{E}[\textcolor{blue}{PS^{-1}P^T}-\bar{P}\bar{P}^T]\succeq 0.
\end{equation}
Taking expectations on both sides of \eqref{equ1} and using \eqref{E1}, \eqref{eqn:lemma5-psd}, and the positive semidefiniteness of
$(\textcolor{blue}{PS^{-1}}-\bar{P})(\textcolor{blue}{PS^{-1}}-\bar{P})^T$,
we obtain
\[
\mathbb{E}[\textcolor{blue}{P_{k|k,\mathrm{ac}}} - \textcolor{blue}{P_{k|k,\mathrm{est}}}] \succeq 0.
\]
Moreover, equality holds if and only if $\mathbb{E}[\textcolor{blue}{P_{k|k-1,\mathrm{est}}}]= P_{k|k-1}$ and $\textcolor{blue}{PS^{-1}}$ is almost surely constant. The latter condition is equivalent to $\textcolor{blue}{K_{k,\text{est}}}=\textcolor{blue}{P_{xz,k|k-1}}\textcolor{blue}{S_{k|k-1}^{-1}}$ being almost surely constant, i.e., unaffected by approximation errors.
\end{pf}

Theorem \ref{theorem1} suggests that using (\ref{oldkf}) to estimate the state covariance matrix tends to lead to overconfident covariance estimation when the estimated $S_{k}$ and $P_{xz,k}$ are not the same as their actual values. Admittedly, in practice, the errors of the approximated covariance matrices $\textcolor{blue}{P_{z,k|k-1}}, \textcolor{blue}{S_{k|k-1}}$, and $\textcolor{blue}{P_{xz,k|k-1}}$ may not be as simple as zero-mean random matrices. However, Theorem \ref{theorem1} does give us a structured explanation as to why nonlinear KFs often underestimate the state covariance matrix. Note that an overconfident state covariance estimation can be very detrimental since it prevents the filter from correcting itself in subsequent iterations, thereby affecting the state estimation accuracy. Again, this problem is unavoidable because the actual values of $S_{k}$ and $P_{xz,k}$ are related to the actual distribution of the states, which is unknown. Therefore, this problem exists for all types of nonlinear KFs as long as (\ref{oldkf}) is used to update the state covariance estimation.

\begin{figure}[htbp]
\centering
\includegraphics[width=8cm]{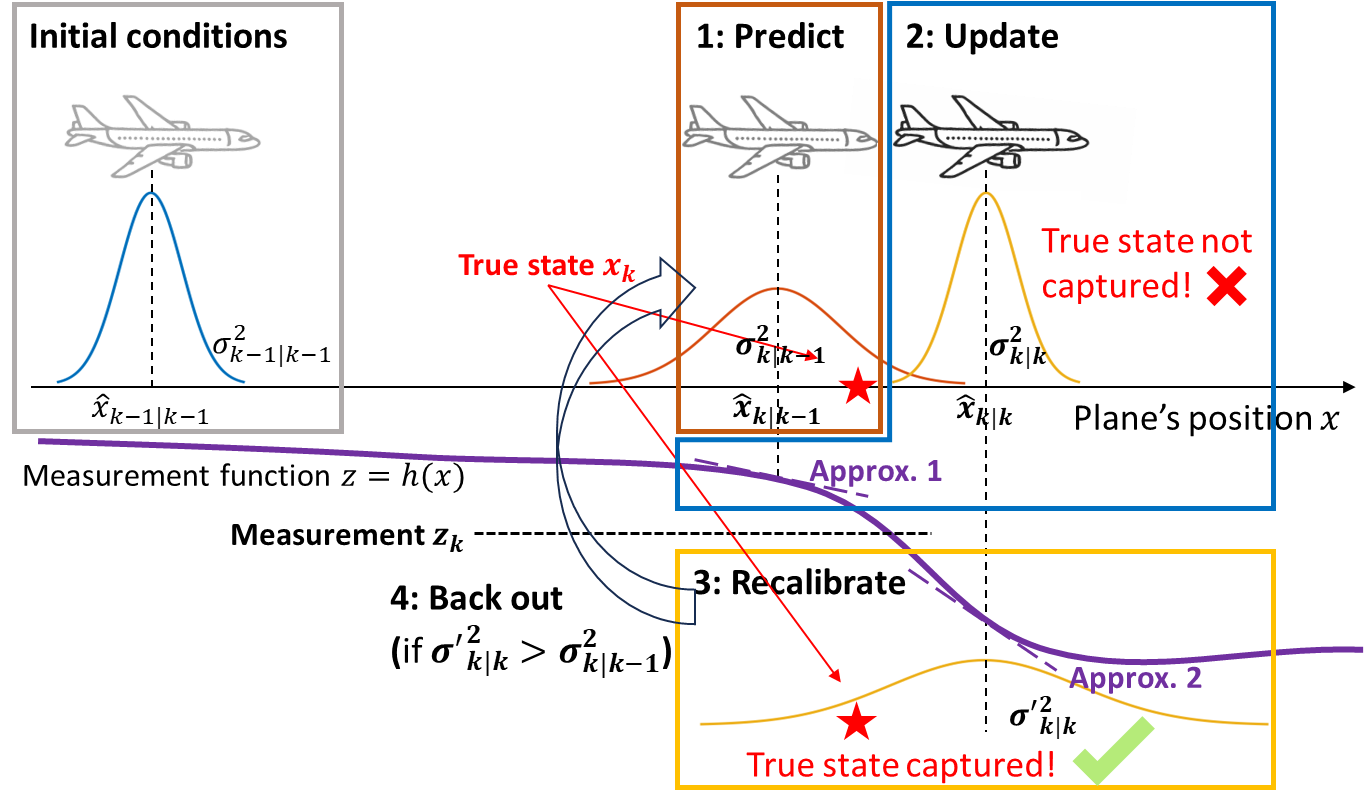}
\caption{Basic idea of the covariance-recalibrated framework for nonlinear KFs. The proposed additional steps re-approximate the innovation covariance and cross-covariance around the updated state, yielding a more accurate state covariance estimate and allowing the update to be withdrawn when it is unhelpful.}
\label{difference}
\end{figure}

In plain language, the problem of the conventional framework originates from the assumption that the selected Kalman gain is optimal. However, the Kalman gain, which was optimized based on the approximated innovation covariance and cross-covariance near the predicted states, generally will not minimize the trace of the actual state covariance. The reason is that the actual states can be far from the predicted states, and the approximations made in the ``predict'' step may be inaccurate. An example of such a problem is shown in Fig. \ref{difference}, where we use an EKF to estimate the position of a plane. In the example, the state is the plane's position along the x-axis, and the measurement function is the altitude of the ground beneath the plane, which is a nonlinear function of the state. Due to the inaccurate approximation of the measurement function around the predicted state (the dashed purple line, ``Approx. 1'', in Fig. \ref{difference}), the EKF overestimates the position of the plane, and the absolute value of the state estimation error $|x_k-\hat{x}_{k|k}|$ becomes higher after the ``update'' step. Meanwhile, the estimated standard deviation of the state, $\sigma_{k|k}$, becomes smaller after the ``update'' step. Such a contradiction suggests that the effectiveness of the Kalman gain is overestimated. This is visualized by the narrow yellow distribution in the ``update'' step in Fig. \ref{difference}, which does not cover the true state. 
\subsection{The proposed covariance-recalibrated framework for nonlinear Kalman filters}
Once we fundamentally understand the problem of the conventional framework, the solution is simple. We can redo the approximation after the update and use the newly approximated innovation covariance and cross-covariance to update the state covariance estimate. In our new framework, this new step is called ``recalibrate''. As shown in Fig.~\ref{difference}, with this additional step, the estimated covariance matrix may not have the smallest possible trace after the update; instead, it will honestly reflect the effect of the Kalman gain on the actual state covariance matrix. This is visualized by the wider yellow distribution in the ``recalibrate'' step, which now covers the true state.

Mathematically, ``recalibrate'' means to  re-estimate the covariance matrices $P_{z,k}$, $S_k$, and $P_{xz,k}$ around the updated states $\hat{x}_{k|k}$. We denote these re-approximated covariance matrices as $\textcolor{blue}{P_{z,k|k}}, \textcolor{blue}{S_{k|k}}$, and $\textcolor{blue}{P_{xz,k|k}}$ since they are all estimated based on the updated states. Similar to (\ref{approx1}), these covariance matrices are calculated by
\begin{equation}\label{approx2}
    \begin{cases} 
    \textcolor{blue}{P_{z,k|k}}: = \Phi_{\mathrm{var}}(h;\hat{x}_{k|k},\textcolor{blue}{P_{k|k-1,\text{est}}}), \\ 
    \textcolor{blue}{S_{k|k}}: = \textcolor{blue}{P_{z,k|k}}+R_k, \\ 
    \textcolor{blue}{P_{xz,k|k}}:= \Phi_{\mathrm{cov}}(h;\hat{x}_{k|k},\textcolor{blue}{P_{k|k-1,\text{est}}}).\end{cases}
\end{equation}

In the new framework, the state covariance matrix is updated by replacing $S_k$ with $\textcolor{blue}{S_{k|k}}$ and $P_{xz,k}$ with $\textcolor{blue}{P_{xz,k|k}}$ in (\ref{P_true}):
\begin{equation}\label{newkf}
\begin{aligned}
    & \textcolor{blue}{P_{k|k,\mathrm{est, new}}}= {\textcolor{blue}{P_{k|k-1,\text{est}}}} + \textcolor{blue}{K_{k,\text{est}}}\textcolor{blue}{S_{k|k}}\textcolor{blue}{K_{k,\text{est}}^T}\\
    &\quad - \textcolor{blue}{P_{xz,k|k}}\textcolor{blue}{K_{k,\text{est}}^T} - \textcolor{blue}{K_{k,\text{est}}}\textcolor{blue}{P_{xz,k|k}^T} \\
    &= {\textcolor{blue}{P_{k|k-1,\text{est}}}} + \textcolor{blue}{P_{xz,k|k-1}}\textcolor{blue}{S_{k|k-1}^{-1}}\textcolor{blue}{S_{k|k}}\textcolor{blue}{S_{k|k-1}^{-1}}\textcolor{blue}{P_{xz,k|k-1}^T} \\
    &\quad - \textcolor{blue}{P_{xz,k|k}}\textcolor{blue}{S_{k|k-1}^{-1}}\textcolor{blue}{P_{xz,k|k-1}^T} - \textcolor{blue}{P_{xz,k|k-1}}\textcolor{blue}{S_{k|k-1}^{-1}}\textcolor{blue}{P_{xz,k|k}^T}.
\end{aligned}
\end{equation}

Note that the proposed covariance-recalibrated framework selects the same Kalman gain and does the same state update as the conventional one; the only difference is that the state covariance update equation is now a direct approximation of (\ref{P_true}), which is the actual state covariance regardless of the selection of the Kalman gain $K$. Intuitively, this can prevent the KF from overconfidence because the equation no longer relies on the problematic assumption that the Kalman gain is optimal. To explain this advantage more rigorously, we consider the newly approximated covariance matrices $\textcolor{blue}{P_{z,k|k}}, \textcolor{blue}{S_{k|k}}$, and $\textcolor{blue}{P_{xz,k|k}}$ produced by a nonlinear KF as
(random) estimators of the actual quantities $P_{z,k}$, $S_k$, and $P_{xz,k}$.
Similar to our previous assumptions about $\textcolor{blue}{P_{z,k|k-1}}, \textcolor{blue}{S_{k|k-1}}$, and $\textcolor{blue}{P_{xz,k|k-1}}$, we regard $\textcolor{blue}{P_{xz,k|k}}$ as a random matrix,
$\textcolor{blue}{P_{z,k|k}}$ as a random positive semidefinite matrix, and
$\textcolor{blue}{S_{k|k}}$ as a random positive definite matrix. Additionally, we assume that these approximated covariance matrices are unbiased estimators and satisfy:
\begin{equation}\label{expcondition3}
    \begin{cases} \mathbb{E}[\textcolor{blue}{P_{z,k|k}}] = P_{z,k} \\ \mathbb{E}[\textcolor{blue}{S_{k|k}}] =S_k \\ \mathbb{E}[\textcolor{blue}{P_{xz,k|k}}]= P_{xz,k}\end{cases} 
\end{equation} 

With these additional definitions and assumptions, we are able to give the following theorem about the advantage of the new framework:
\begin{thm}\label{theorem3}
Assume that conditions \eqref{expcondition} and \eqref{expcondition3} hold. 
In addition, assume that
\[
(\textcolor{blue}{P_{xz,k|k-1}},\, \textcolor{blue}{S_{k|k-1}})
\quad\text{and}\quad
(\textcolor{blue}{P_{xz,k|k}},\, \textcolor{blue}{S_{k|k}})
\]
are independent, and that $\mathbb{E}\!\left[\textcolor{blue}{P_{k|k-1,\mathrm{est}}}\right]=P_{k|k-1}.$
Then the estimator \(\textcolor{blue}{P_{k|k,\mathrm{est,new}}}\) defined by \eqref{newkf} is an unbiased estimator of \(\textcolor{blue}{P_{k|k,\mathrm{ac}}}\), i.e.,
\begin{equation}
\mathbb{E}\!\left[\textcolor{blue}{P_{k|k,\mathrm{est,new}}}-\textcolor{blue}{P_{k|k,\mathrm{ac}}}\right]
=0.
\end{equation}
\end{thm}
\begin{pf}
    With the assumptions made in Theorem \ref{theorem3}, we have:
    \begin{equation}
    \begin{aligned}
    &\mathbb{E}[\textcolor{blue}{P_{k|k,\mathrm{est, new}}}-\textcolor{blue}{P_{k|k,\mathrm{ac}}}]\\
    =&\mathbb{E}[\textcolor{blue}{P_{xz,k|k-1}}\textcolor{blue}{S_{k|k-1}^{-1}}(\textcolor{blue}{S_{k|k}}-S_k)\textcolor{blue}{S_{k|k-1}^{-1}}\textcolor{blue}{P_{xz,k|k-1}^T}]\\
    &-\mathbb{E}[(\textcolor{blue}{P_{xz,k|k}}-P_{xz,k})\textcolor{blue}{S_{k|k-1}^{-1}}\textcolor{blue}{P_{xz,k|k-1}^T}]\\
    &-\mathbb{E}[\textcolor{blue}{P_{xz,k|k-1}}\textcolor{blue}{S_{k|k-1}^{-1}}(\textcolor{blue}{P_{xz,k|k}}-P_{xz,k})^T]\\
    =&\,0.
    \end{aligned}
    \end{equation}
The last equality holds because the terms in parentheses have zero expectation, due to the unbiasedness (i.e., $\mathbb{E}[\textcolor{blue}{S_{k|k}}]=S_{k}$ and $\mathbb{E}[\textcolor{blue}{P_{xz,k|k}}]=P_{xz,k}$) and independence assumptions. 
\end{pf}

Theorem \ref{theorem3} suggests that the proposed covariance-recalibrated framework can generally prevent the filter from being overconfident in the ``update'' step if the approximation errors of the measurement function at the predicted and updated states are independent of each other. Admittedly, this assumption about the independence of the approximation errors is very hard to verify and may not hold in practice. Nonetheless, the theorem does explain why recalibrating the covariance matrices $P_{z,k}, S_k$, and $P_{xz,k}$ after the state update is advantageous for accurate state covariance estimation. Specifically, the advantage comes from our replacement of (\ref{oldkf}) with (\ref{newkf}), which highlights the fundamental idea of the new framework. That is, instead of reducing the updated state covariance matrix to the smallest possible value, we re-approximate the system after the state update to estimate the actual effect of the Kalman gain more accurately. 

In nonlinear KFs, the update can sometimes make the state estimate less accurate and increase the actual state covariance, as shown in Fig. \ref{difference}. If this situation were detectable, a natural reaction would be to withdraw the update and use the more accurate predicted state (and the predicted state covariance) as the final output. Unfortunately, such detection is impossible based solely on the estimated state covariance matrix in the conventional KF framework, since $\textcolor{blue}{P_{k|k,\mathrm{est}}}$ calculated in (\ref{oldkf}) is guaranteed to satisfy $\textcolor{blue}{P_{k|k-1,\mathrm{est}}} - \textcolor{blue}{P_{k|k,\mathrm{est}}} \succeq 0$. In the new framework, however, the additional “recalibrate” step makes such detection possible, which motivates us to introduce a ``back out'' step. The ``back out'' step uses the reported covariance trace to select between the predicted and updated state estimates. Specifically, by the end of the recalibrate step, the filter has two candidate estimates of the current state: the predicted state $\hat{x}_{k|k-1}$ with covariance $\textcolor{blue}{P_{k|k-1,\mathrm{est}}}$, and the updated state $\hat{x}_{k|k}$ with covariance $\textcolor{blue}{P_{k|k,\mathrm{est,new}}}$. We retain the updated estimate unless its reported covariance has a larger trace than the predicted covariance. In that case, we return the predicted state and covariance. If the ``back out'' step is omitted, the filter may produce a suboptimal estimate. In addition, the recalibrated covariance $\textcolor{blue}{P_{k|k,\mathrm{est,new}}}$ is no longer guaranteed to satisfy $\textcolor{blue}{P_{k|k-1,\mathrm{est}}} - \textcolor{blue}{P_{k|k,\mathrm{est,new}}} \succeq 0$. Therefore, its trace can in some cases be substantially larger than that of $\textcolor{blue}{P_{k|k-1,\mathrm{est}}}$, which motivates the ``back out'' step.

The pseudo-code of the covariance-recalibrated framework is shown in Algorithm \ref{new_algorithm}. Note that the new framework degrades to the conventional framework when the ``recalibrate'' and ``back out'' steps are replaced with (\ref{oldkf}). Moreover, when all measurement functions are linear, the approximations of $P_{z,k}$, $S_k$, and $P_{xz,k}$ become identical before and after the recalibration step. In this case, the proposed framework is exactly equivalent to the conventional framework and preserves its optimality for linear systems.
\begin{algorithm}
	\caption{The proposed covariance-recalibrated framework for nonlinear Kalman filters}\label{new_algorithm}
	\begin{algorithmic}[1]
        \Statex \textbf{Input:} Process noise covariance matrix $\boldsymbol{Q}_k$, Measurement noise covariance matrix $\boldsymbol{R}_k$, state transition function $\boldsymbol{f}(\boldsymbol{x},\boldsymbol{u})$, measurement function $\boldsymbol{h}{(\boldsymbol{x})}$, system inputs $\boldsymbol{u}_k$, measurements $\boldsymbol{z}_k$
        \Statex \textbf{Initialization:}
        \State $\boldsymbol{\hat{x}}_{0|0}=\mathbb{E}[\boldsymbol{x}_0]$
        \State $\boldsymbol{P}_{0|0}=\mathbb{E}[(\boldsymbol{\hat{x}}_{0|0}-\boldsymbol{x}_0)(\boldsymbol{\hat{x}}_{0|0}-\boldsymbol{x}_0)^T]$
		\For {every time step $k$}
        \Statex \hspace{1em} \textbf{Predict:}
        \State Estimate $\boldsymbol{\hat{x}}_{k|k-1}$ and $\boldsymbol{P}_{k|k-1}$
        \Statex \hspace{1em} (based on $\boldsymbol{\hat{x}}_{k-1|k-1}, \boldsymbol{P}_{k-1|k-1}, \boldsymbol{f}, \boldsymbol{u}_{k-1},$ and $\boldsymbol{Q}_{k-1}$)
		\Statex \hspace{1em} \textbf{Update:}
        \State Estimate $\boldsymbol{\hat{z}}_{k|k-1}, \boldsymbol{P}_{xz,k|k-1}$ and $\boldsymbol{P}_{z,k|k-1}$ 
        \Statex \hspace{1em} (based on $\boldsymbol{\hat{x}}_{k|k-1}, \boldsymbol{P}_{k|k-1},$ and $\boldsymbol{h}$)
        \State $\boldsymbol{S}_{k|k-1}=\boldsymbol{P}_{z,k|k-1}+\boldsymbol{R}_k$
        \State $\boldsymbol{K}_k=\boldsymbol{P}_{xz,k|k-1}\boldsymbol{S}_{k|k-1}^{-1}$
        \State $\boldsymbol{\hat{x}}_{k|k}=\boldsymbol{\hat{x}}_{k|k-1}+\boldsymbol{K}_k(\boldsymbol{z}_k-\boldsymbol{\hat{z}}_{k|k-1})$
		\Statex \hspace{1em} \textbf{Recalibrate:}
        \State Estimate $\boldsymbol{P}_{xz,k|k}$ and $\boldsymbol{P}_{z,k|k}$
        \Statex \hspace{1em} (based on $\boldsymbol{\hat{x}}_{k|k}, \boldsymbol{P}_{k|k-1},$ and $ \boldsymbol{h}$)
        \State $\boldsymbol{S}_{k|k}=\boldsymbol{P}_{z,k|k}+\boldsymbol{R}_k$
        \scriptsize \State {$\boldsymbol{P}_{k|k}=\boldsymbol{P}_{k|k-1}+\boldsymbol{K}_k\boldsymbol{S}_{k|k}\boldsymbol{K}_k^T-\boldsymbol{P}_{xz,k|k}\boldsymbol{K}_k^T-\boldsymbol{K}_k\boldsymbol{P}_{xz,k|k}^T$}
        \normalsize \Statex \hspace{1em} \textbf{Back out:}
        \If{$\text{tr}(\boldsymbol{P}_{k|k})>\text{tr}(\boldsymbol{P}_{k|k-1})$}
        \State $\boldsymbol{\hat{x}}_{k|k}=\boldsymbol{\hat{x}}_{k|k-1}$
        \State $\boldsymbol{P}_{k|k}=\boldsymbol{P}_{k|k-1}$
        \EndIf
		\EndFor
        \Statex where $\boldsymbol{\hat{x}}$ is the estimated state, $\boldsymbol{P}$ is the states' covariance matrix, $\boldsymbol{\hat{z}}$ is the estimated measurement, $\boldsymbol{P}_{xz}$ is the covariance between the states and the measurements, $\boldsymbol{P}_{z}$ is the covariance of the estimated measurements, $\boldsymbol{S}$ is the innovation (or residual) covariance, and $\boldsymbol{K}_k$ is the Kalman gain. The subscript $k|k-1$ represents the estimated value before the state update, and $k|k$ represents the estimated value after the state update.
	\end{algorithmic}
\end{algorithm}

\begin{figure}[htbp]
\centering
\includegraphics[width=7.2cm]{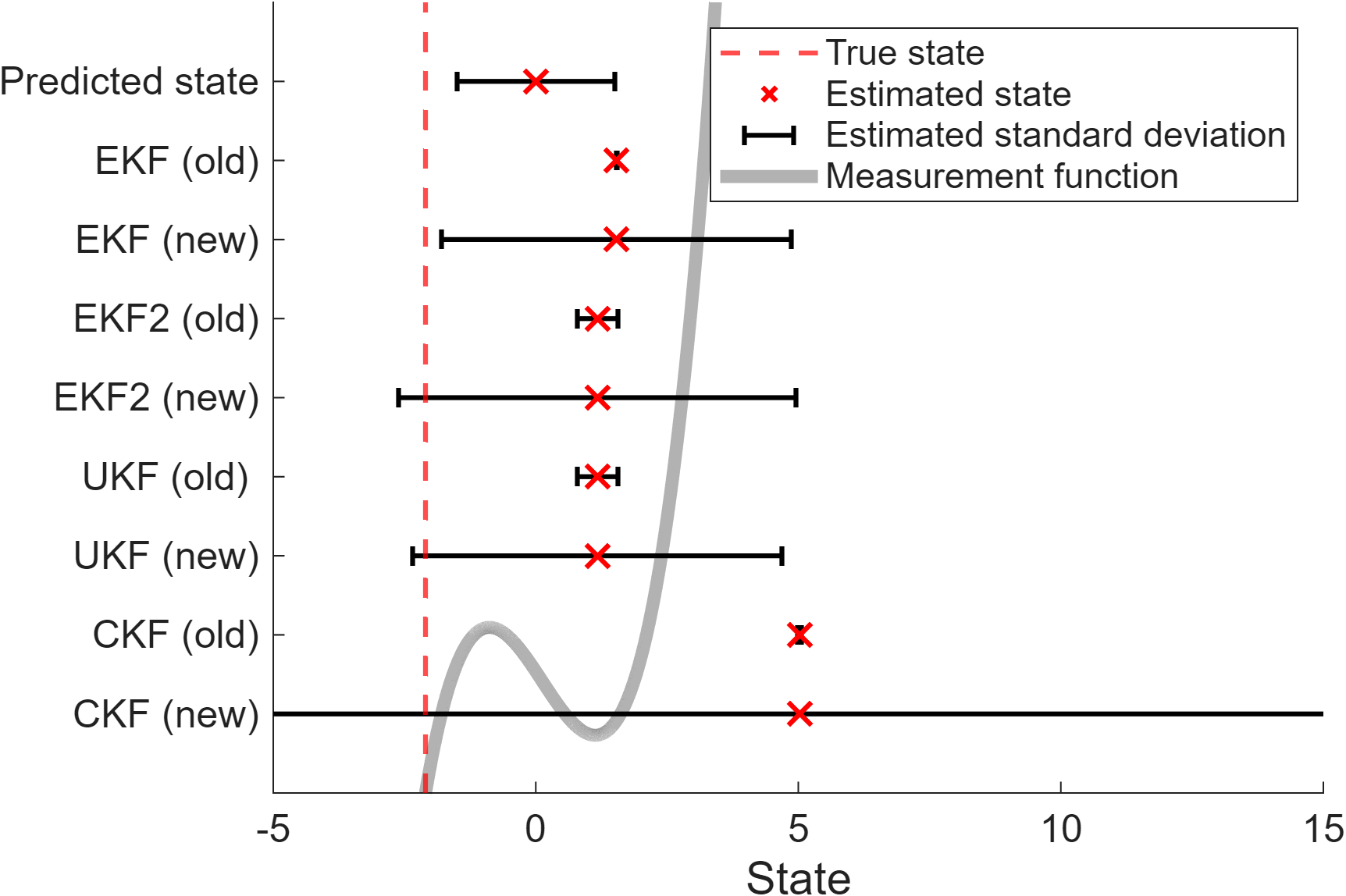}
\caption{The comparison of the variance estimation after the state update and recalibration using different frameworks and different types of nonlinear Kalman filters. The recalibrated variance of CKF (new) is approximately $144^2$, almost 10,000 times larger than the predicted variance, which underscores the importance of the ``back out'' step.}
\label{update_comparison}
\end{figure}

To provide an intuitive illustration of the roles of the ``recalibrate'' and ``back out'' steps, Fig.~\ref{update_comparison} shows a single update for several nonlinear KFs applied to the scalar measurement model
\begin{equation}
    z = h(x) + v,\qquad h(x) = x^3/3 - x^2/8 - x + 1.5383,
\end{equation}
with the observed measurement $z_1 = 0$. The measurement function $h(x)$ is plotted as the thick gray curve. The predicted state is $\hat{x}_{1|0} = 0$ with variance $1.5^2$, and the measurement noise standard deviation is $\sigma_z = 0.01$. This example is particularly challenging because $h'(\hat{x}_{1|0}) < 0$ while the true state lies to the left of $\hat{x}_{1|0}$, so a local linearization can easily mislead the filter. The state and covariance estimates produced by different algorithms are shown in separate rows in the figure. Under the conventional framework, all KF variants substantially underestimate the actual estimation error, as indicated by the overly narrow $\pm 1\sigma$ intervals that fail to cover the true state. In contrast, within the proposed framework, the covariance estimates are much more consistent with the true error. In this example, the recalibrated variance correctly indicates that the update is less reliable than the prediction, so the ``back out'' step is triggered and the filter discards the misleading update, outputting the predicted state and its variance as the final estimate. If the ``back out'' step is removed from the framework, the estimated state becomes less accurate for all filters. Moreover, in this example, the CKF's estimated variance after recalibration is approximately $144^2$, which is almost 10{,}000 times the variance of the predicted state. Such a large increase in the estimated covariance further illustrates the practical value of the ``back out'' step.

\subsection{Relationship to Iterated Kalman Filter}\label{relation_IEKF}
The covariance-recalibrated framework described above may remind the readers of the iterated Kalman filter (IKF), which also re-approximates the measurement function after calculating the Kalman gain \cite{IKF, IUKF, ICKF}. However, IKF is fundamentally different from the covariance-recalibrated framework in three ways. Firstly, IKF redoes the approximation and recalculates the Kalman gain iteratively until convergence. However, such a convergence is not guaranteed, so the algorithm is less robust and often requires fine-tuning the step size of each iteration \cite{IKF_update}. Even if the iteration converges, the convergence may take a very long time. By contrast, the covariance-recalibrated framework requires no iteration, and the re-approximation is only done once. Secondly, IKF was initially only used on EKF, and the derivation of iterated EKF (IEKF) requires an explicit approximation of the measurement functions, which is only possible for EKF and EKF2. Although a few attempts have been made to extend IKF to UKF and CKF \cite{IUKF, ICKF, IUKF2, IUKF3}, a consensus on how to do this robustly has not been reached. Therefore, it is not easy to implement IKF on all types of nonlinear KFs while guaranteeing that the estimator can converge in most cases. Our framework, on the other hand, does not have this problem since there is no convergence issue. Thirdly, and most importantly, the basic idea behind IKF is still to find the optimal Kalman gain and not to estimate the covariance matrix accurately. Namely, IKF still updates the covariance matrix using (\ref{oldkf}), so the covariance-underestimation result of Theorem~\ref{theorem1} also applies when its assumptions hold. When the algorithm cannot converge properly, the estimation can be even worse because the estimated $P_{z,k}, S_k$, and $P_{xz,k}$ can be less accurate, which can lead to a higher degree of overconfidence. In general, in terms of making a more accurate estimation, finding the optimal Kalman gain is not as important as making the covariance estimation accurate, as we will show later in the simulation results.

Since IKF also uses (\ref{oldkf}) to update the state covariance matrix, it retains the covariance update of the conventional framework. A natural question is whether adding covariance recalibration can improve its estimation accuracy. For the IEKF considered here, however, recalibration becomes redundant at convergence. The algorithm in the supplementary material linearizes the measurement function at $x_{i-1}$ (rows 9--13). At convergence, $x_{i-1}\approx x_i=\hat{x}_{k|k}$, so repeating the moment calculation at the updated estimate with the same predicted covariance leaves the covariance report essentially unchanged. The proposed framework instead recalibrates the covariance after a single state update, without iterating the gain. In the numerical examples, this approach is more effective: Figs.~\ref{exp_3D}--\ref{exp_battery} show that the EKF combined with the proposed framework is consistently more accurate than the IEKF, while also requiring less computation.

\section{Results}

To demonstrate the effectiveness of the proposed framework, we validate it in five different applications suitable for nonlinear KFs. Table \ref{setup_brief} briefly describes the five applications. The system models and parameter setup are detailed in Appendix C. For each scenario, we apply both the conventional and proposed frameworks to four nonlinear KFs: EKF, EKF2, UKF, and CKF. We also implement the IEKF for comparison. For reference, Appendix~\ref{appendix_algo} presents a representative EKF implementation of the proposed framework, whereas the full set of algorithm listings used in the experiments—including the conventional IEKF benchmark and the proposed implementations of EKF2, CKF, and UKF—is provided as supplementary material in the public repository. All nine $(4+4+1)$ nonlinear KFs are used for state estimation under different measurement-noise setups (the covariances of the measurement noise differ across setups). To ensure a fair comparison and reduce the impact of Monte Carlo randomness, each KF is simulated 10,000 times for each measurement-noise level. All filters share the same realizations of the random initial-state errors, process noise, and measurement noise. Concretely, these random sequences are generated from the same random seed so that the $i^{\text{th}}$ simulation of each method (e.g., EKF, UKF, CKF) is driven by identical noise realizations.

\begin{table*}[htbp]\scriptsize
\centering
\caption{A brief description of the five applications of nonlinear Kalman filters investigated in this paper.}\label{setup_brief}
\begin{tabular}{|c|c|c|c|c|c|}
\hline
Applications & \begin{tabular}[c]{@{}c@{}}3D target\\ tracking\end{tabular} & \begin{tabular}[c]{@{}c@{}}Terrain-referenced\\ navigation\end{tabular} & \begin{tabular}[c]{@{}c@{}}Synchronous generator\\ state estimation\end{tabular} & \begin{tabular}[c]{@{}c@{}}Pendulum state\\ estimation\end{tabular} & \begin{tabular}[c]{@{}c@{}}Battery state\\ estimation\end{tabular} \\ \hline
Number of states & 6 & 2 & 4 & 2 & 3 \\ \hline
Linear state transition function? & Yes & Yes & No & No & No \\ \hline
Number of measurements & 2 & 1 & 1 & 1 & 1 \\ \hline
Linear measurement functions? & No & No & No & No & No \\ \hline
Number of inputs & 3 & 2 & 3 & 1 & 1 \\ \hline
Number of iterations & 30 & 100 & 100 & 100 & 180 \\ \hline
\end{tabular}
\end{table*}

\subsection{State estimation accuracy}
\begin{figure}[htbp]
\centering
\includegraphics[width=7.2cm]{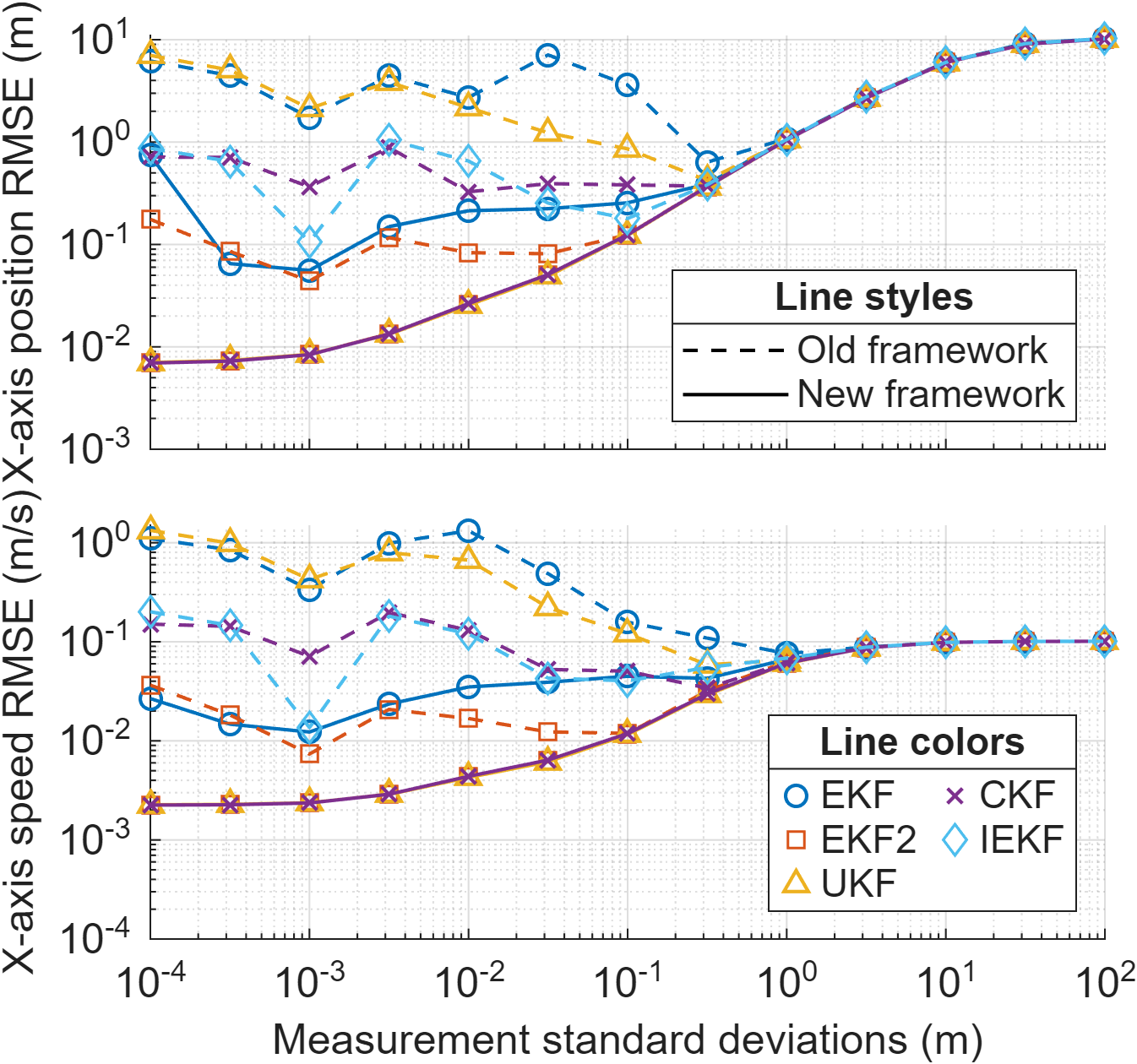}
\caption{The root mean squared error of the state estimations under different measurement noise setups (3D target tracking).}
\label{exp_3D}
\end{figure}

\begin{figure}[htbp]
\centering
\includegraphics[width=7.2cm]{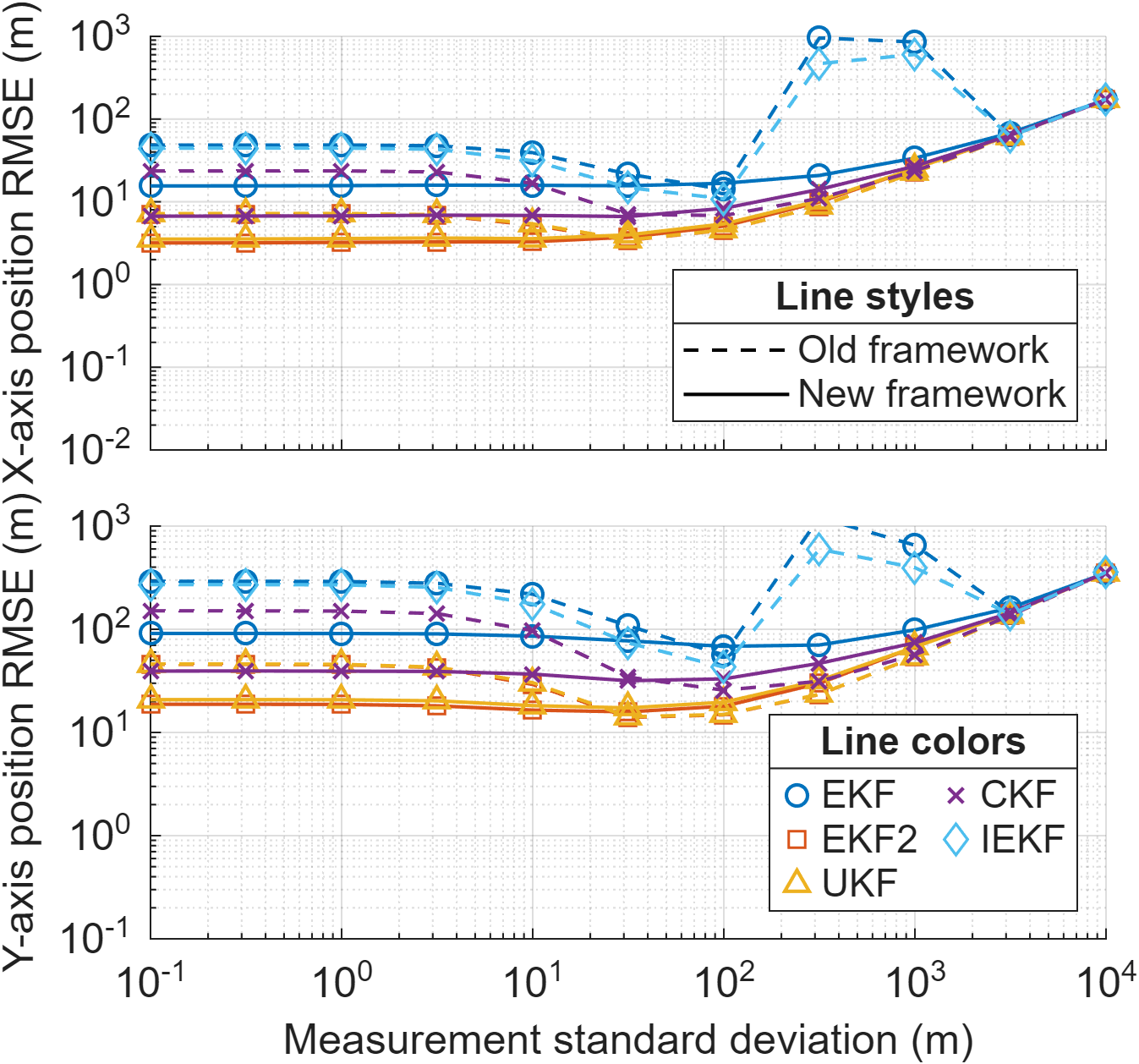}
\caption{The root mean squared error of the state estimations under different measurement noise setups (terrain-referenced navigation).}
\label{exp_navigation}
\end{figure}

\begin{figure}[htbp]
\centering
\includegraphics[width=7.2cm]{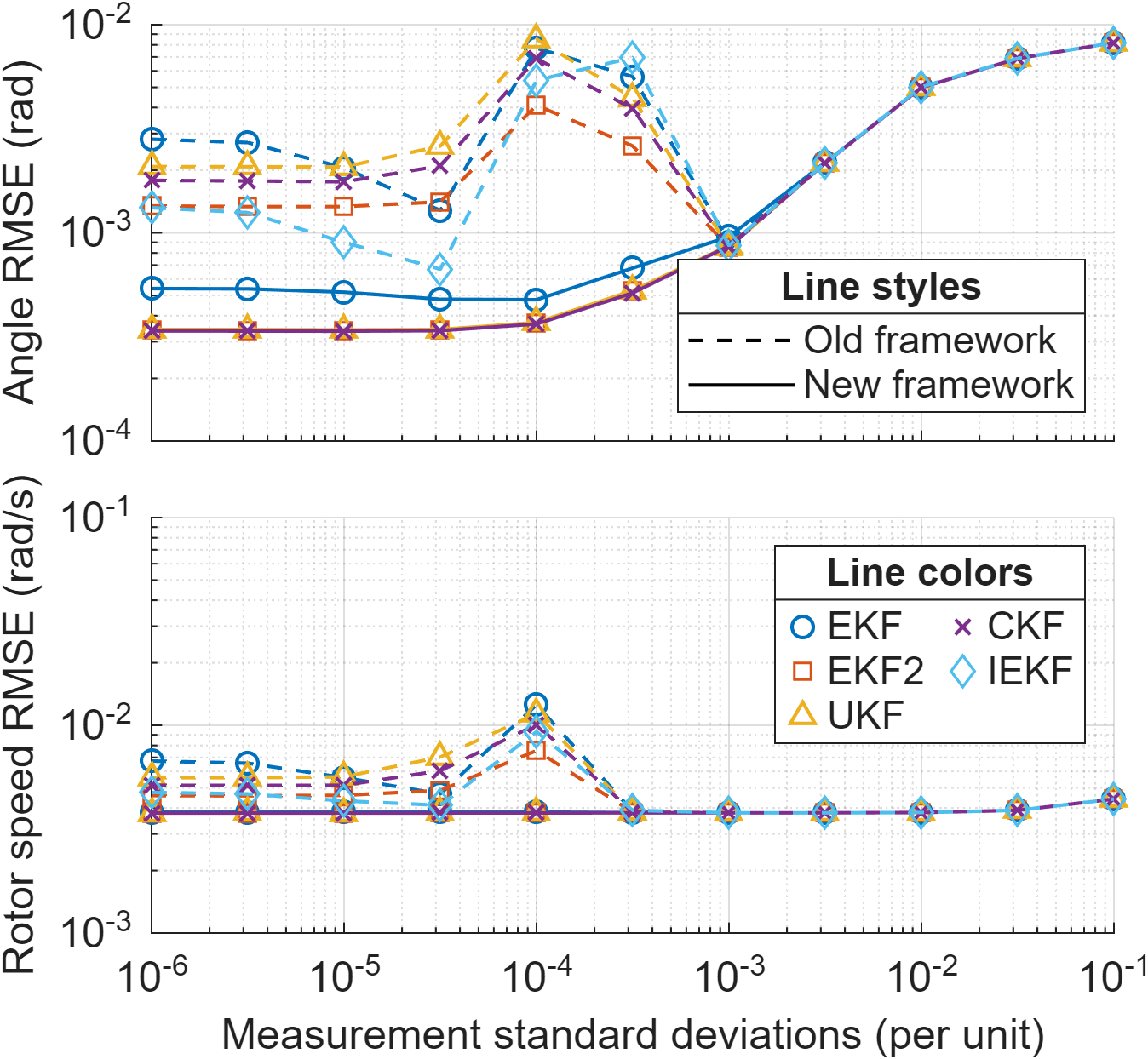}
\caption{The root mean squared error of the state estimations under different measurement noise setups (synchronous generator state estimation).}
\label{exp_generator}
\end{figure}

\begin{figure}[htbp]
\centering
\includegraphics[width=7.2cm]{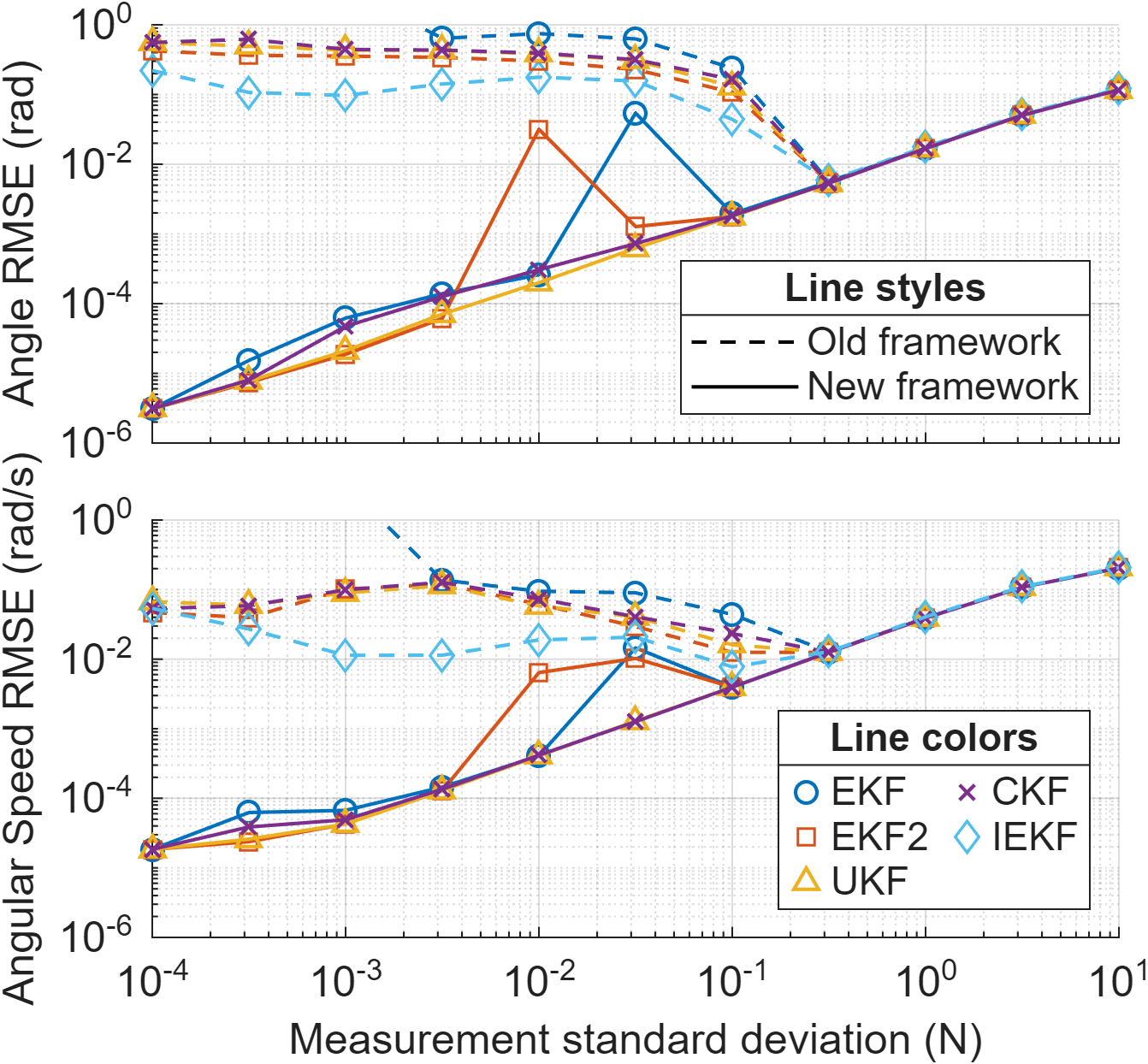}
\caption{The root mean squared error of the state estimations under different measurement noise setups (pendulum state estimation).}
\label{exp_pendulum}
\end{figure}

\begin{figure}[htbp]
\centering
\includegraphics[width=7.2cm]{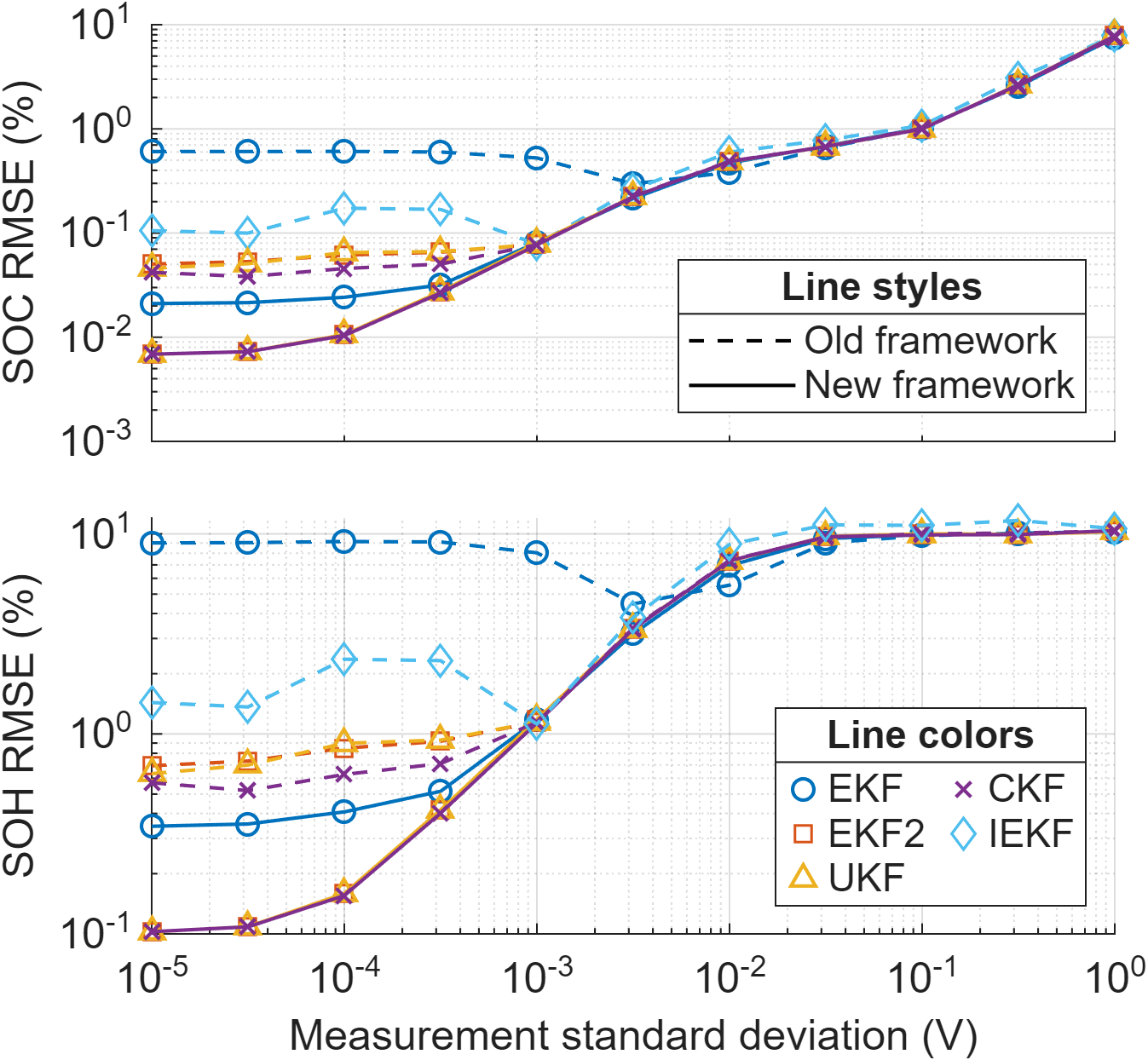}
\caption{The root mean squared error of the state estimations under different measurement noise setups (battery state estimation).}
\label{exp_battery}
\end{figure}

The root mean squared error (RMSE) of the state estimations given by different nonlinear KFs is shown in Figs. \ref{exp_3D}–\ref{exp_battery}. In all figures, different types of nonlinear KFs are represented by different colors and markers, and the old and new frameworks are represented by dashed lines and solid lines, respectively. The x-axis represents the standard deviations of the measurements. The root mean squared error is calculated from the estimation errors in 10,000 simulations. The estimation error is calculated as the difference between the estimated state and the true state at the end of the final iteration of each simulation. Note that we only show the RMSE of two representative states for systems with more than two states. For example, in 3D target tracking, the six states are the object’s speed and position along the three axes, and Fig. \ref{exp_3D} examines the object’s speed and position estimation accuracy along the x-axis. As can be seen from the five figures, the simulation results are very similar to each other. Therefore, we mainly analyze the first application, 3D target tracking, in depth in this section.

As shown in Figs. \ref{exp_3D}–\ref{exp_battery}, the proposed framework (solid lines) significantly improves the accuracy of all types of nonlinear KFs. The improvement becomes stronger as the measurement noise becomes smaller. This result aligns with our expectations because the proposed framework only changes how the measurements are used in the nonlinear KF, and the estimations will rely more on the measurements instead of system knowledge as the measurement noise decreases. For all these applications, when the standard deviations of the measurements are small, the RMSE of all four types of nonlinear KFs can usually be reduced by more than an order of magnitude, which is a significant improvement.

Another interesting observation of Figs. \ref{exp_3D}–\ref{exp_battery} is that the RMSE of EKF2, UKF, and CKF becomes quite similar after the new framework is used. This phenomenon can be explained by the fact that all three methods consider the first and second-order terms in the Taylor expansion of the nonlinear state transition and measurement functions in certain ways \cite{relation}. On the other hand, EKF is generally slightly worse than the other three types of nonlinear KFs because it only considers the first-order terms in the Taylor expansion.

By comparing the performance of IEKF and EKF (with and without the proposed framework) in Figs. \ref{exp_3D}–\ref{exp_battery}, we observe that although the IEKF outperforms the EKF under the conventional framework, it is consistently inferior to the EKF implemented with the new framework, while also incurring higher computational cost. As discussed in Section~\ref{relation_IEKF}, this discrepancy arises because the IEKF continues to produce overconfident covariance estimates, particularly in cases where the iteration fails to converge — a situation that occurs frequently in practice.

\begin{figure}[htbp]
\centering
\includegraphics[width=7.2cm]{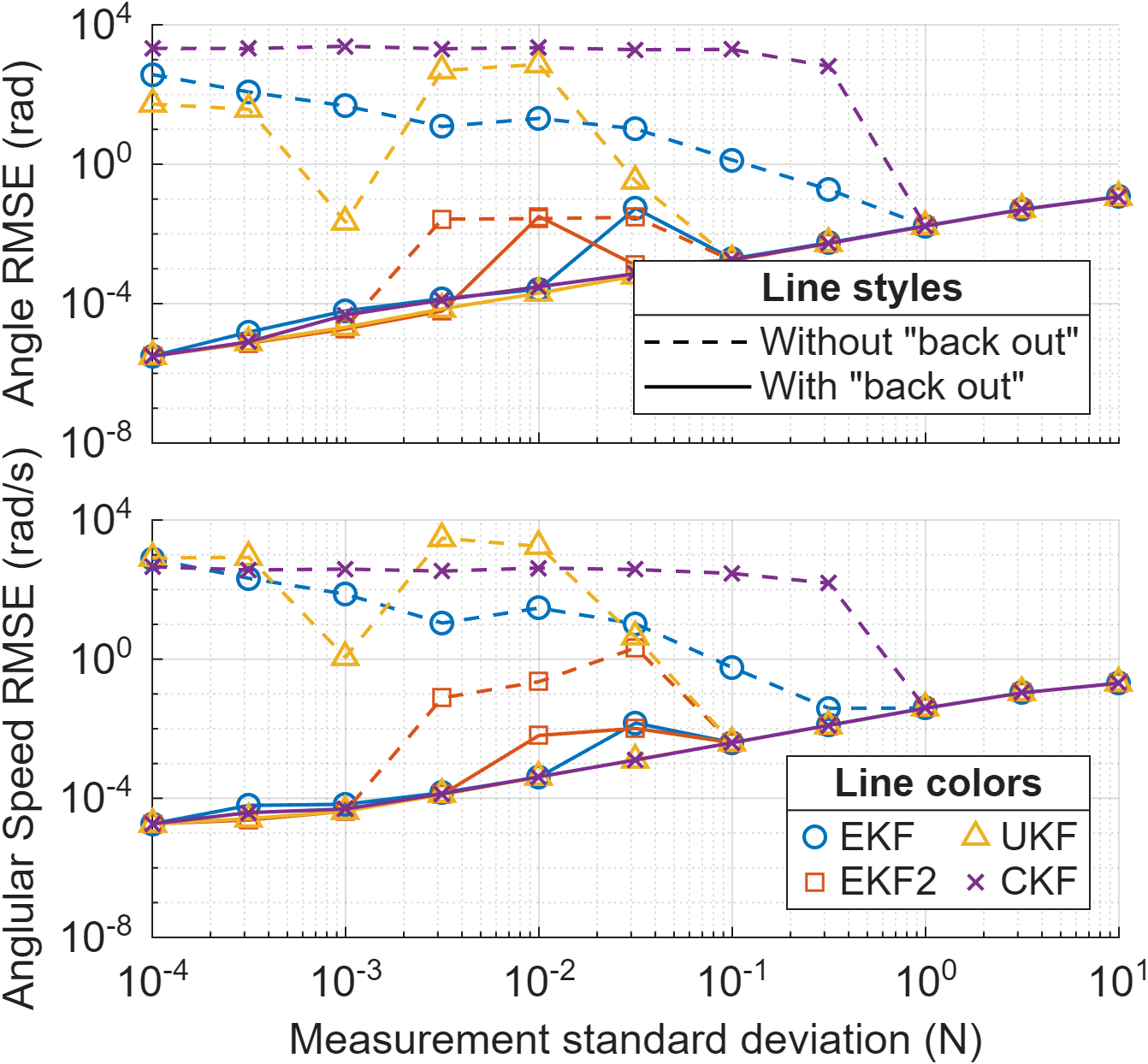}
\caption{The root mean squared error of the state estimations with and without the ``back out'' step under different measurement noise setups (pendulum state estimation).}
\label{backout}
\end{figure}

\begin{figure}[htbp]
\centering
\includegraphics[width=7.2cm]{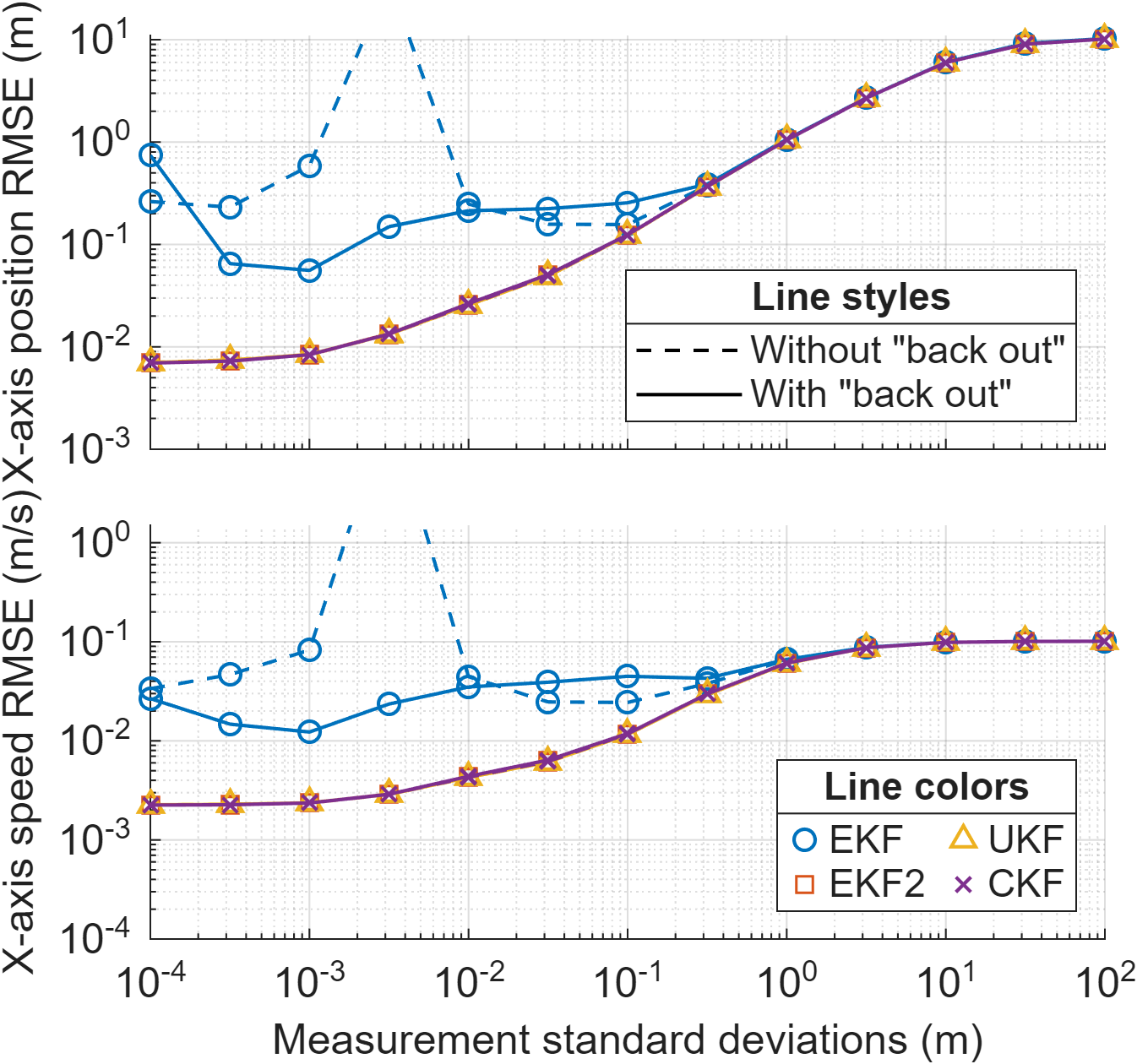}
\caption{The root mean squared error of the state estimations with and without the ``back out'' step under different measurement noise setups (3D target tracking).}
\label{backout2}
\end{figure}

\begin{figure}[htbp]
\centering
\includegraphics[width=7.2cm]{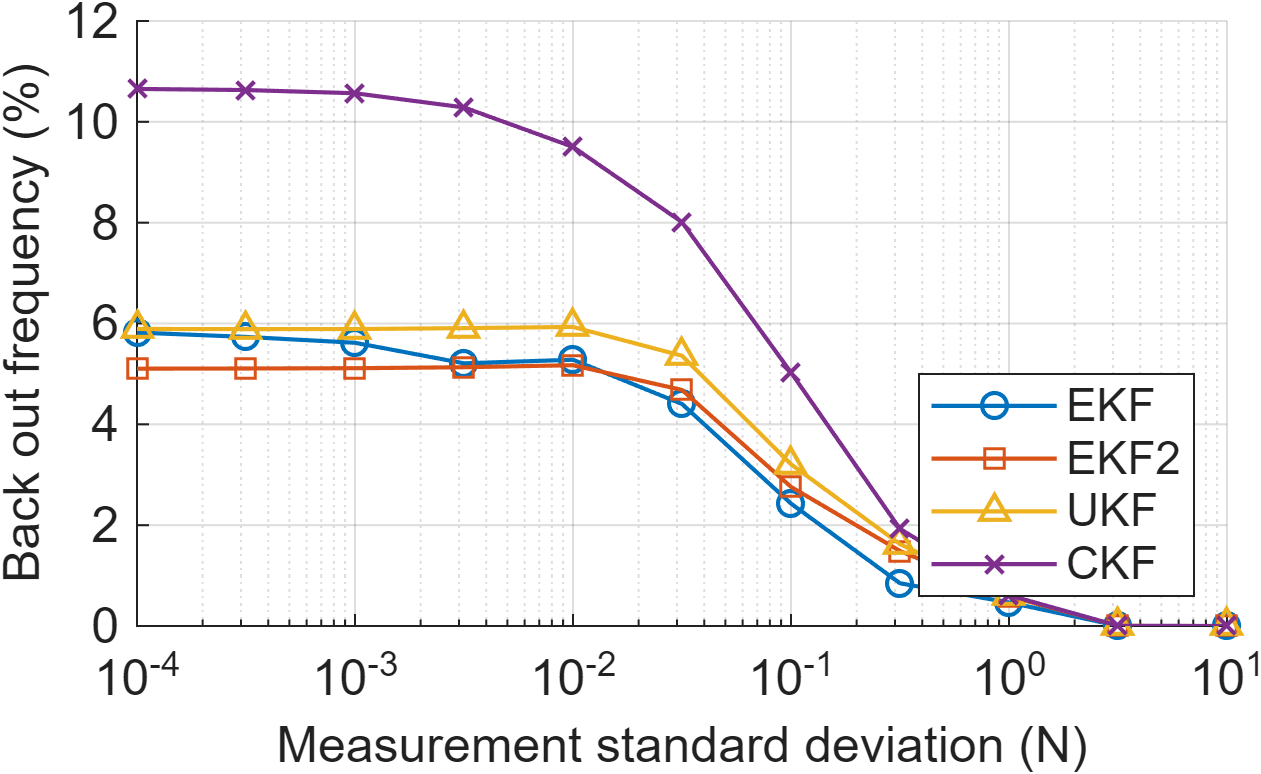}
\caption{Percentage of state updates that trigger ``back out'' (pendulum state estimation).}
\label{A4}
\end{figure}

The practical value of the ``back out'' step is illustrated in Fig.~\ref{backout}, which compares the RMSE of pendulum state estimation (Application 4) with and without this step. The results show that, without the ``back out'' step, all four nonlinear KF variants become significantly less accurate when the measurement noise is low. This degradation is related to the fact that the a posteriori covariance computed in (\ref{newkf}) is no longer guaranteed to be smaller than the predicted covariance in the positive semidefinite ordering. As illustrated previously in Fig.~\ref{update_comparison}, the recalibrated covariance can increase substantially after the update. Without the ``back out'' step, the filter may retain an inferior updated estimate instead of reverting to the predicted one, thereby reducing estimation accuracy in subsequent iterations. It should be noted, however, that the effect of the ``back out'' step is application dependent. For example, as shown in Fig.~\ref{backout2}, removing this step in Application~1 noticeably degrades the performance of the EKF, whereas the EKF2, UKF, and CKF are almost unaffected. Nevertheless, it is still advisable to retain this step, since it helps prevent excessive growth of the estimated covariance while incurring negligible computational overhead in the proposed framework. To provide a quantitative sense of how often this safeguard is activated in practice, Fig.~\ref{A4} reports the percentage of state updates that trigger ``back out'' in the pendulum state estimation example. The corresponding results for the other applications are qualitatively similar; additional plots are provided in the supplementary material available in the public repository.

\subsection{Covariance estimation accuracy}
\begin{figure}[htbp]
\centering
\includegraphics[width=7.2cm]{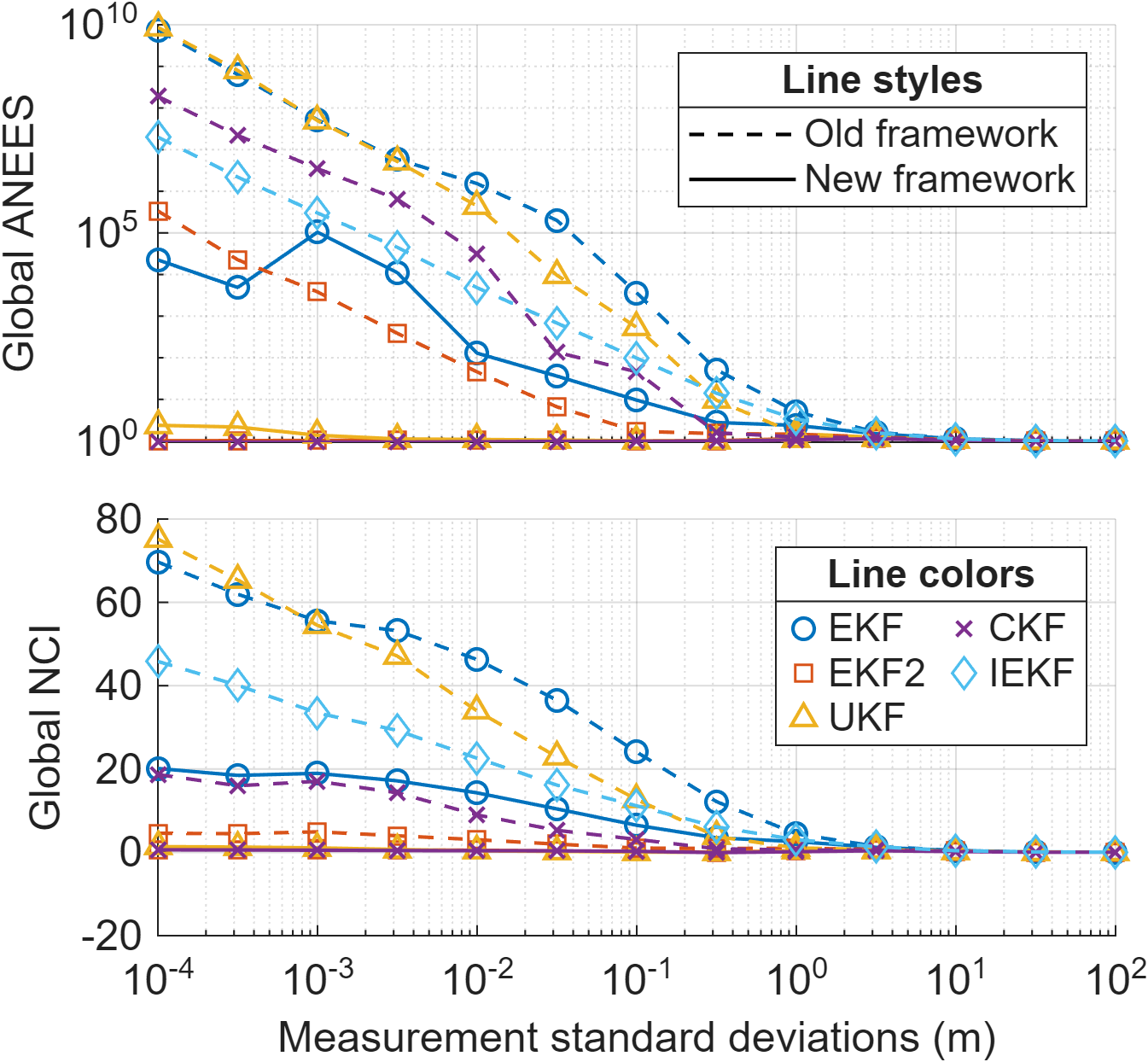}
\caption{The global Average Normalized Estimation Error Squared (ANEES) and Noncredibility Index (NCI) of various nonlinear Kalman filters under different measurement noise setups. The covariance-recalibrated framework significantly mitigates the problem of overconfidence.}
\label{variance}
\end{figure}

As discussed previously, the proposed framework aims to improve covariance estimation by re-approximating the measurement functions after each state update. To evaluate whether the framework actually leads to more accurate covariance estimates, we introduce two metrics for assessing covariance estimation accuracy: the Average Normalized Estimation Error Squared (ANEES) and the Noncredibility Index (NCI). Specifically, the ANEES at time step $k$ is defined as \cite{ANEES}:
\scriptsize\begin{equation}
    \mathrm{ANEES}_k:=
    \frac{1}{n_x N_\text{MC}}
    \sum_{i=1}^{N_\text{MC}}
    \bigl(x_k^{(i)}-\hat{x}_{k|k}^{(i)}\bigr)^T
    \bigl(\textcolor{blue}{P^{(i)}_{k,\mathrm{est}}}\bigr)^{-1}
    \bigl(x_k^{(i)}-\hat{x}_{k|k}^{(i)}\bigr),
\end{equation}\normalsize
where $n_x$ is the state dimension and $N_\text{MC}=10{,}000$ is the number of Monte Carlo simulations. 
The superscript $(i)$ denotes the $i^\text{th}$ Monte Carlo simulation. 
Here, $x_k^{(i)}$ and $\hat{x}_{k|k}^{(i)}$ are the true and estimated states at time step $k$, respectively, and 
$\textcolor{blue}{P^{(i)}_{k,\mathrm{est}}}$ is the corresponding state estimation error covariance matrix. 
Note that $\textcolor{blue}{P^{(i)}_{k,\mathrm{est}}} = \textcolor{blue}{P^{(i)}_{k|k,\mathrm{est}}}$ for the conventional framework, 
whereas $\textcolor{blue}{P^{(i)}_{k,\mathrm{est}}} = \textcolor{blue}{P^{(i)}_{k|k,\mathrm{est, new}}}$ for the proposed framework. When $\textcolor{blue}{P^{(i)}_{k,\mathrm{est}}}$ equals the true error covariance and the estimation error is zero-mean, we have $\mathbb{E}[\mathrm{ANEES}_k]=1$. Therefore, the deviation of $\mathrm{ANEES}_k$ from~1 reflects the degree of overconfidence or conservatism in the KF. However, since the estimation error can be biased in practice, the value of ANEES alone may not be a sufficiently reliable indicator of overconfidence \cite{ANEES}. Therefore, we also consider the metric NCI, which quantifies the mismatch between the estimated covariance and the empirical second-moment matrix of the estimation error. Specifically, the NCI at time step $k$ is defined as \cite{NCI}:
\scriptsize
\begin{equation}
    \mathrm{NCI}_k:=
    \frac{10}{N_\text{MC}}
    \sum_{i=1}^{N_\text{MC}}
    \log_{10}
    \frac{
    \bigl(x_k^{(i)}-\hat{x}_{k|k}^{(i)}\bigr)^T
    \bigl(\textcolor{blue}{P^{(i)}_{k,\mathrm{est}}}\bigr)^{-1}
    \bigl(x_k^{(i)}-\hat{x}_{k|k}^{(i)}\bigr)
    }{
    \bigl(x_k^{(i)}-\hat{x}_{k|k}^{(i)}\bigr)^T
    \bigl(P^*_{k}\bigr)^{-1}
    \bigl(x_k^{(i)}-\hat{x}_{k|k}^{(i)}\bigr)
    },
\end{equation}
\normalsize
where $P^*_{k}$ is the sample error second-moment matrix at time step $k$, defined by
\begin{equation}
    P^*_{k}:=
    \frac{1}{N_\text{MC}}
    \sum_{i=1}^{N_\text{MC}}
    \bigl(x_k^{(i)}-\hat{x}_{k|k}^{(i)}\bigr)
    \bigl(x_k^{(i)}-\hat{x}_{k|k}^{(i)}\bigr)^T.
\end{equation}
By construction, $\mathrm{NCI}_k \approx 0$ when $\textcolor{blue}{P^{(i)}_{k,\mathrm{est}}}$ matches $P_k^*$, $\mathrm{NCI}_k > 0$ indicates overconfidence, and $\mathrm{NCI}_k < 0$ indicates conservatism. Given the definitions of ANEES and NCI for each step $k$, we define the global ANEES and NCI as the average across all time steps:
\begin{equation}
    \begin{cases}
    \mathrm{ANEES}=\dfrac{1}{N_{\text{step}}}\displaystyle\sum_{k=1}^{N_{\text{step}}}\mathrm{ANEES}_k,\\[0.8em]
    \mathrm{NCI}=\dfrac{1}{N_{\text{step}}}\displaystyle\sum_{k=1}^{N_{\text{step}}}\mathrm{NCI}_k,
    \end{cases}
\end{equation}
where $N_{\text{step}}$ is the total number of time steps in the simulation. Given these metrics, Fig.~\ref{variance} compares the global ANEES and NCI of various KF algorithms using the conventional and proposed frameworks in the target tracking application. Under the conventional framework, when the standard deviations of both measurements are below 0.1\,m, all nonlinear KFs substantially underestimate the covariance, regardless of the specific algorithm. In contrast, under the proposed framework, only the EKF remains noticeably overconfident, while the ANEES and NCI of the other nonlinear KFs are very close to one and zero, respectively. This residual gap for the EKF is consistent with its neglect of second- and higher-order terms in the measurement functions; nevertheless, its covariance estimates are still significantly more accurate when combined with the proposed framework.

\subsection{Convergence characteristics}
\begin{figure}[htbp]
\centering
\includegraphics[width=7.2cm]{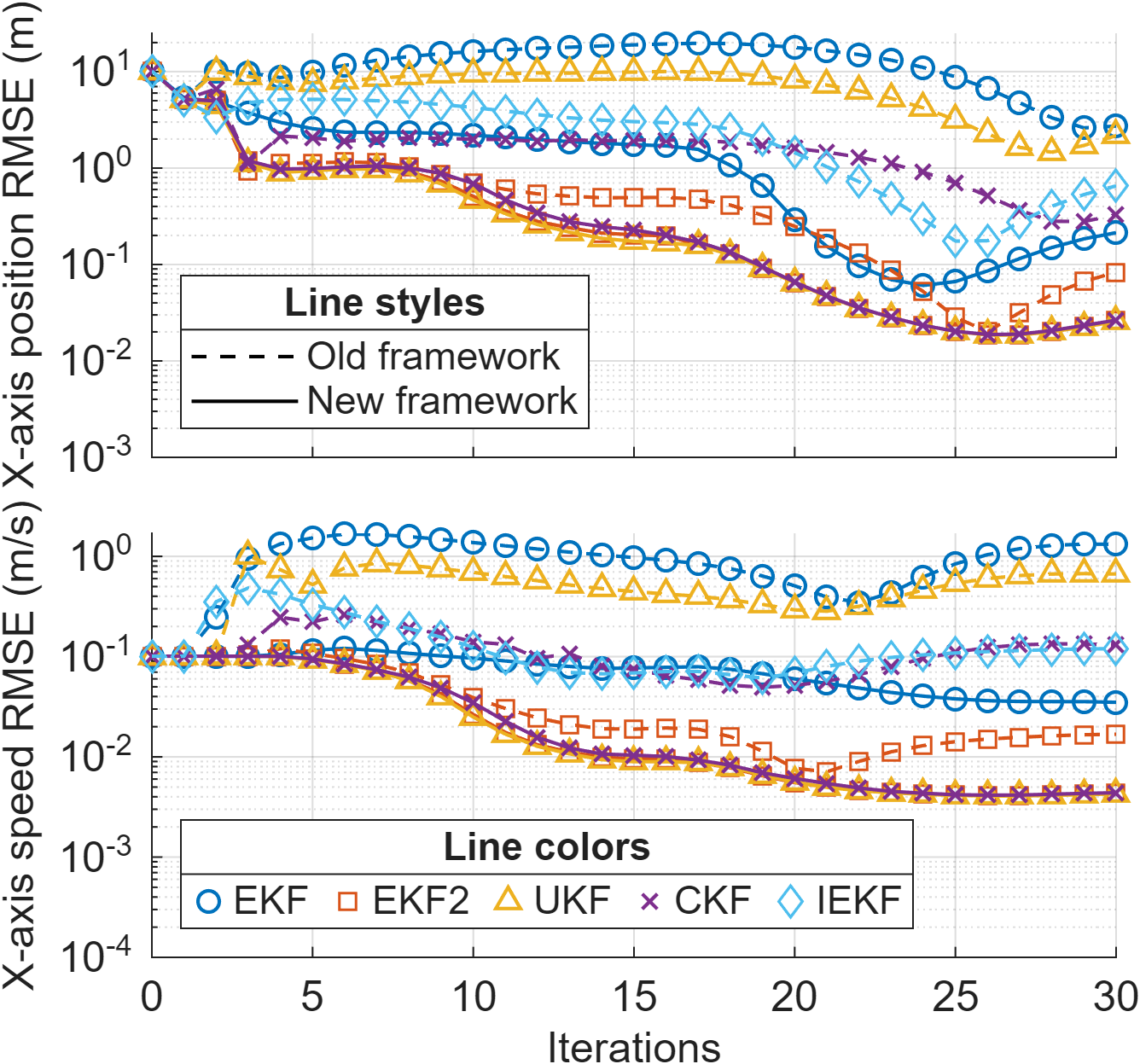}
\caption{The root mean squared error of the state estimations after different numbers of iterations.}
\label{convergence}
\end{figure}

Another important metric for evaluating the performance of different filters is the convergence characteristics. We visualize this metric by fixing the standard deviations of the two measurements to 0.01 m and plotting the RMSE of the state estimations after different numbers of iterations. The results are shown in Fig. \ref{convergence}. From the figure, we can see that all types of nonlinear KFs can converge faster after being combined with the new framework. This result suggests that the proposed framework is very suitable for applications that require high convergence speed.  

\subsection{Computational time}
\begin{figure}[htbp]
\centering
\includegraphics[width=8cm]{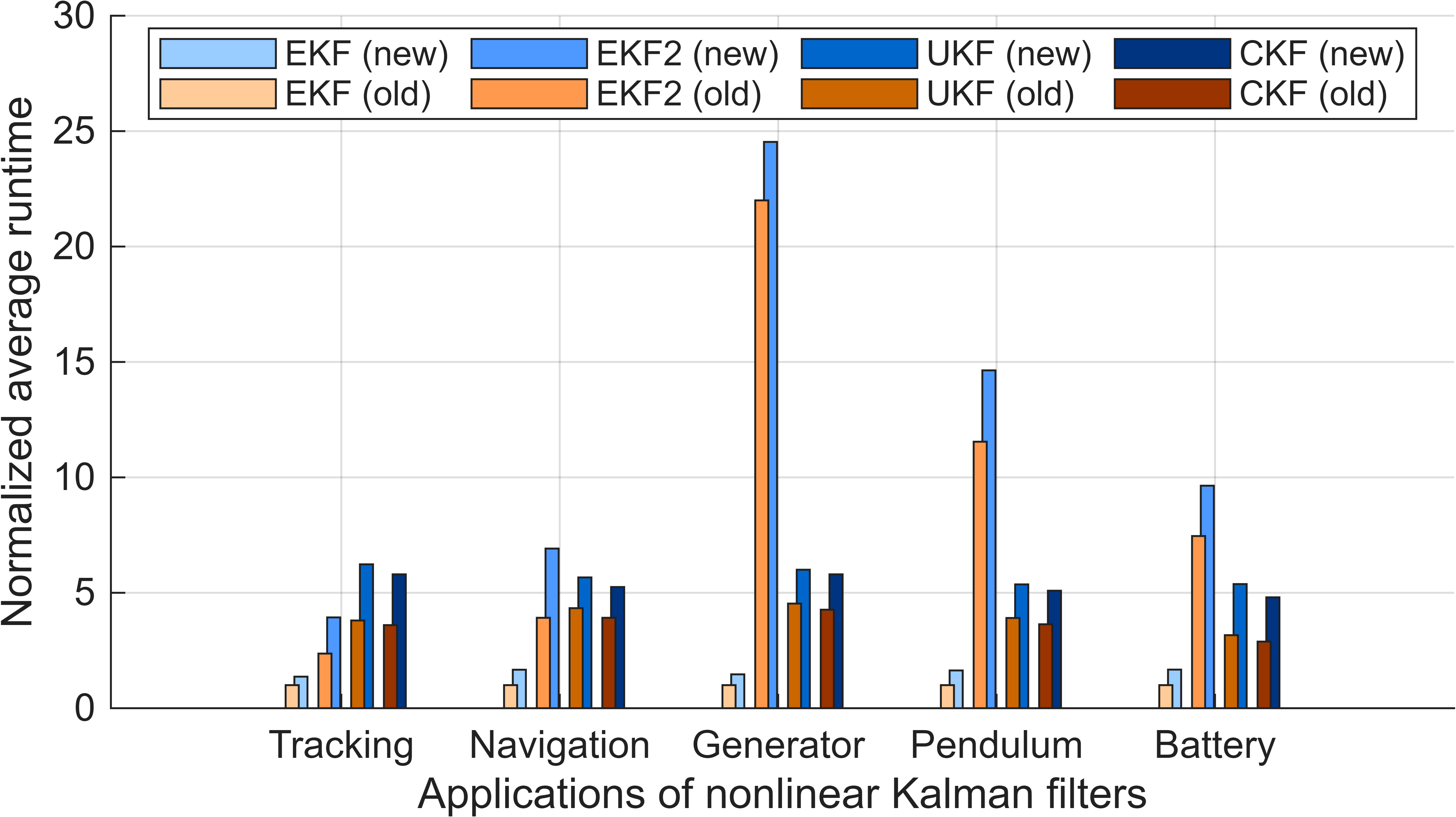}
\caption{Comparison of the normalized computational time of different methods and frameworks.}
\label{bar}
\end{figure}

Since the extra steps introduced in the covariance-recalibrated framework take time, it is also important to compare the computational time of the two frameworks. The results are summarized in Fig. \ref{bar}. Note that the average runtime shown in the figure is calculated from over 100,000 simulations run on the same computer. Specifically, for each combination of nonlinear KF and framework, we recorded the method's runtime whenever it was used and calculated the average. After the average runtime is calculated, the value is normalized with respect to the average runtime of the conventional EKF for the same application. In other words, the normalized average runtime of ``EKF (old)'' always equals one in Fig. \ref{bar}. 

From the result, we can see that the new framework generally increases the computational time by 10–90\%, varying for different nonlinear systems. Namely, when the system has linear state transition functions and nonlinear measurement functions (which is true for the first two applications in the table), the ``update'' step should account for most of the computational time in the conventional framework and re-approximating the measurement functions can almost double the runtime in the worst case. However, noticing that the new framework can reduce the estimation error by more than an order of magnitude when the measurement noise is small, we can say that the extra computational complexity is generally worth it. Another thing to note is that the combination of EKF and the new framework requires significantly less computational time than the combination of any other nonlinear KFs and the old framework. At the same time, the former’s accuracy is generally higher than the latter in the applications we examined in this paper. Therefore, even in situations where the computational time is of major concern, the proposed framework can still be helpful because the new EKF can outperform the old EKF2, UKF, and CKF, in addition to saving a lot of computational time.

\section{Conclusion}
In this paper, we propose a new covariance-recalibrated framework applicable to almost all types of nonlinear KFs. The new framework re-approximates the innovation covariance and cross-covariance after the state update, yielding a more accurate covariance update. The framework largely mitigates the problem of overconfident covariance estimation, an inherent limitation of conventional nonlinear KFs. The new framework is tested in five different applications, reducing estimation errors by more than an order of magnitude when measurement noise is low and significantly shortening the convergence time. Future work will focus on the stability analysis of the proposed framework and on combining it with deliberate enlargement of the effective process and measurement covariance, which may further improve filter stability and accuracy in strongly nonlinear settings.

\begin{ack}                               
This material is based upon work supported, in part, by the National Science Foundation under Grant No. 1847177.  
\end{ack}

\bibliographystyle{unsrt}        
\bibliography{autosam}           



\appendix
\section{Supplementary proof for Theorem \ref{theorem1}}\label{appendix_proof}

\begin{lem}\label{lemmakey}
    Let $\textcolor{blue}{S}$ be a random symmetric and positive definite matrix whose mean is the identity matrix $I$. Let $\bm{v}$ be a random vector whose mean is $\bm{v}_0$. The following inequality holds: 
    \begin{equation}\label{eq:lemma_bound}
        \mathbb{E}(\bm{v}^T\textcolor{blue}{S^{-1}}\bm{v})\geq \bm{v}_0^T\bm{v}_0.
    \end{equation}
    Note that the equality sign only holds when $\textcolor{blue}{S^{-1}}\bm{v}$ is constant almost surely.
\end{lem}

\begin{pf}
    If $\bm{v}_0=0$, \eqref{eq:lemma_bound} follows from positive definiteness of $\textcolor{blue}{S}$, with equality only when $\bm{v}=0$ almost surely. Hence assume $\bm{v}_0\ne0$.
    Let $\bm{x}:=\textcolor{blue}{S^{-\frac{1}{2}}}\bm{v},\bm{y}:=\textcolor{blue}{S^{\frac{1}{2}}}\bm{v}_0$. Since $\bm{v}$ is a random vector with a mean of $\bm{v}_0$, and $\textcolor{blue}{S}$ is a random symmetric matrix with a mean of the identity matrix,
\begin{equation}
    \begin{cases}
    \mathbb{E}[\bm{x}^T\bm{x}]=\mathbb{E}[\bm{v}^T\textcolor{blue}{S^{-1}}\bm{v}]\\
    \mathbb{E}[\bm{y}^T\bm{y}]=\mathbb{E}[\bm{v}_0^T\textcolor{blue}{S}\bm{v}_0]=\bm{v}_0^T\mathbb{E}[\textcolor{blue}{S}]\bm{v}_0=\bm{v}_0^T\bm{v}_0\\
    \mathbb{E}[\bm{x}^T\bm{y}]=\mathbb{E}[\bm{v}^T\bm{v}_0]=\mathbb{E}[\bm{v}]^T\bm{v}_0=\bm{v}_0^T\bm{v}_0
    \end{cases} 
\end{equation}
    Note that $\mathbb{E}[\bm{x}^T\bm{y}]$ is an inner product on $L^2(\Omega;\mathbb{R}^n)$ (see Chapter 2.1 in \cite{inner1} or Chapter 6.8 in \cite{inner2}). Applying Cauchy--Schwarz inequality, we have
    \begin{equation}
        \mathbb{E}[\bm{v}^T\textcolor{blue}{S^{-1}}\bm{v}]\geq \frac{(\bm{v}_0^T\bm{v}_0)^2}{\bm{v}_0^T\bm{v}_0},
    \end{equation}
    which completes the proof. Note that the equality holds only when $\textcolor{blue}{S^{-1}}\bm{v}$ is constant almost surely.
\end{pf}

\section{The detailed algorithm of nonlinear Kalman filters used in the paper}\label{appendix_algo}
A representative implementation of the proposed framework using EKF is shown in Algorithm~\ref{EKF}. Note that for the EKF specifically, step 18 can be written equivalently as the Joseph form $\boldsymbol{P}_{k|k}=(\boldsymbol{I}-\boldsymbol{K}_k\boldsymbol{H}_{k|k})\boldsymbol{P}_{k|k-1}(\boldsymbol{I}-\boldsymbol{K}_k\boldsymbol{H}_{k|k})^T+\boldsymbol{K}_k\boldsymbol{R}_k\boldsymbol{K}_k^T$, which follows by substituting steps 15--17 into step 18. The Joseph form expresses the covariance as a sum of positive semidefinite terms and improves numerical robustness. The corresponding implementations using EKF2, CKF, and UKF, together with the conventional IEKF benchmark used for comparison, are provided in the supplementary material available in the public repository. Replacing the ``Recalibrate'' and ``Back out'' steps in the proposed framework with (13) recovers the conventional framework.

\begin{algorithm}
	\caption{The extended Kalman filter with the new framework}\label{EKF}
		\begin{algorithmic}[1]
        \Statex \textbf{Input:} Process noise covariance matrix $\boldsymbol{Q}_k$, Measurement noise covariance matrix $\boldsymbol{R}_k$, state transition function $\boldsymbol{f}(\boldsymbol{x},\boldsymbol{u})$, measurement function $\boldsymbol{h}{(\boldsymbol{x})}$, system inputs $\boldsymbol{u}_k$, measurements $\boldsymbol{z}_k$
        \Statex \textbf{Initialization:}
        \State $\boldsymbol{\hat{x}}_{0|0}=\mathbb{E}[\boldsymbol{x}_0]$
        \State $\boldsymbol{P}_{0|0}=\mathbb{E}[(\boldsymbol{\hat{x}}_{0|0}-\boldsymbol{x}_0)(\boldsymbol{\hat{x}}_{0|0}-\boldsymbol{x}_0)^T]$
		\For {every time step $k$}
        \Statex \hspace{1em} \textbf{Predict:}
        \State $\boldsymbol{F}_k=\frac{\partial \boldsymbol{f}}{\partial \boldsymbol{x}} |_{\boldsymbol{\hat{x}}_{k-1|k-1},\boldsymbol{u}_{k-1}}$
        \State $\boldsymbol{\hat{x}}_{k|k-1}=\boldsymbol{f}(\boldsymbol{\hat{x}}_{k-1|k-1},\boldsymbol{u}_{k-1})$
        \State $\boldsymbol{P}_{k|k-1}=\boldsymbol{F}_k\boldsymbol{P}_{k-1|k-1}\boldsymbol{F}_k^T+\boldsymbol{Q}_k$ 
		\Statex \hspace{1em} \textbf{Update:}
        \State $\boldsymbol{H}_{k|k-1}= \frac{\partial \boldsymbol{h}}{\partial \boldsymbol{x}} |_{\boldsymbol{\hat{x}}_{k|k-1}}$
        \State $\boldsymbol{\hat{z}}_{k|k-1} = \boldsymbol{h}(\boldsymbol{\hat{x}}_{k|k-1})$
        \State $\boldsymbol{P}_{xz,k|k-1}=\boldsymbol{P}_{k|k-1}\boldsymbol{H}_{k|k-1}^T$
        \State $\boldsymbol{P}_{z,k|k-1}=\boldsymbol{H}_{k|k-1}\boldsymbol{P}_{k|k-1}\boldsymbol{H}_{k|k-1}^T$
        \State $\boldsymbol{S}_{k|k-1}=\boldsymbol{P}_{z,k|k-1}+\boldsymbol{R}_k$
        \State $\boldsymbol{K}_k=\boldsymbol{P}_{xz,k|k-1}\boldsymbol{S}_{k|k-1}^{-1}$
        \State $\boldsymbol{\hat{x}}_{k|k}=\boldsymbol{\hat{x}}_{k|k-1}+\boldsymbol{K}_k(\boldsymbol{z}_k-\boldsymbol{\hat{z}}_{k|k-1})$
		\Statex \hspace{1em} \textbf{Recalibrate:}
        \State $\boldsymbol{H}_{k|k}= \frac{\partial \boldsymbol{h}}{\partial \boldsymbol{x}} |_{\boldsymbol{\hat{x}}_{k|k}}$
        \State $\boldsymbol{P}_{xz,k|k}=\boldsymbol{P}_{k|k-1}\boldsymbol{H}_{k|k}^T$
        \State $\boldsymbol{P}_{z,k|k}=\boldsymbol{H}_{k|k}\boldsymbol{P}_{k|k-1}\boldsymbol{H}_{k|k}^T$
        \State $\boldsymbol{S}_{k|k}=\boldsymbol{P}_{z,k|k}+\boldsymbol{R}_k$
        \scriptsize\State {$\boldsymbol{P}_{k|k}=\boldsymbol{P}_{k|k-1}+\boldsymbol{K}_k\boldsymbol{S}_{k|k}\boldsymbol{K}_k^T-\boldsymbol{P}_{xz,k|k}\boldsymbol{K}_k^T-\boldsymbol{K}_k\boldsymbol{P}_{xz,k|k}^T$}
        \normalsize \Statex \hspace{1em} \textbf{Back out:}
        \If{$\text{tr}(\boldsymbol{P}_{k|k})>\text{tr}(\boldsymbol{P}_{k|k-1})$}
        \normalsize \State $\boldsymbol{\hat{x}}_{k|k}=\boldsymbol{\hat{x}}_{k|k-1}$
        \State $\boldsymbol{P}_{k|k}=\boldsymbol{P}_{k|k-1}$
        \EndIf
		\EndFor
	\end{algorithmic} 
\end{algorithm}

\section{The detailed simulation setup} \label{appendix_setup}
This paper uses five different applications of nonlinear KFs to test the effectiveness of the proposed covariance-recalibrated framework: 3D target tracking, terrain-referenced navigation, synchronous generator state estimation, pendulum state estimation, and battery state estimation.
\begin{figure}[htbp]
\centering
\includegraphics[width=8cm]{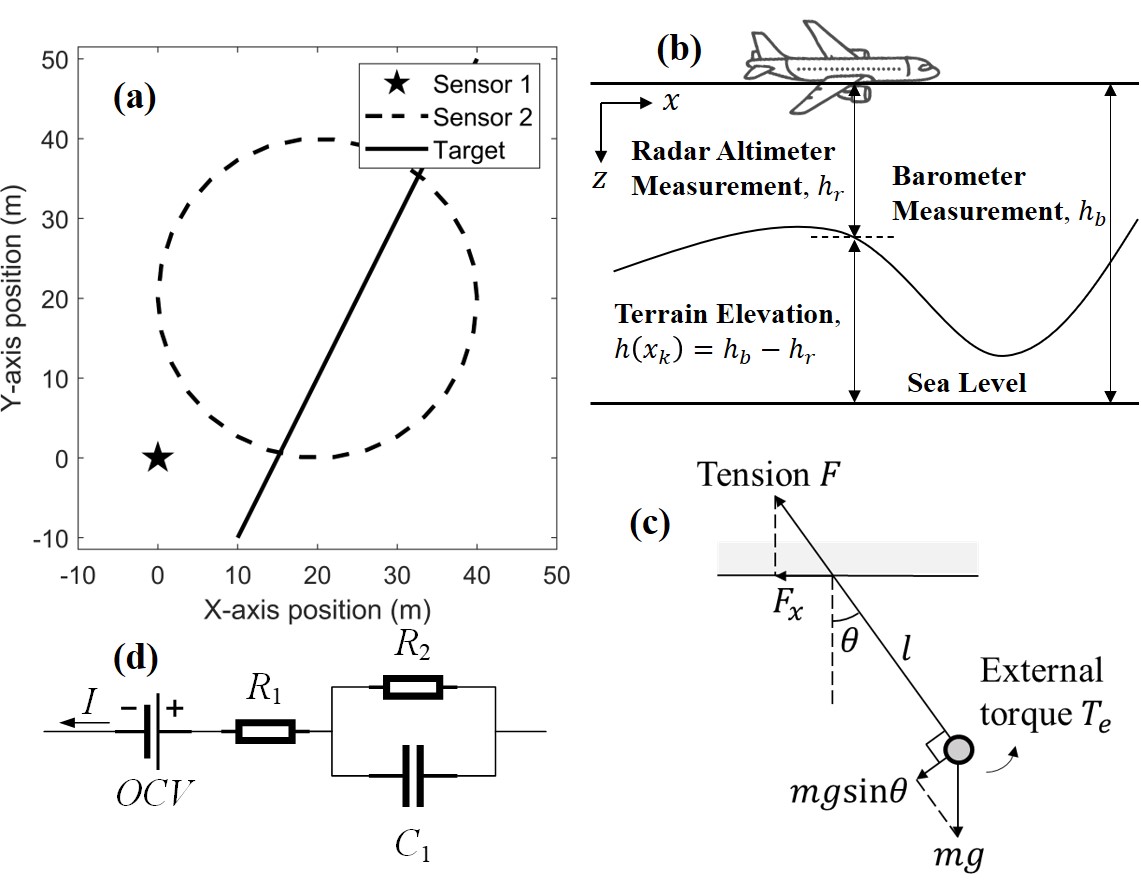}
\caption{Examples of nonlinear systems used in this paper for validation. (a) 3D object tracking. (b) Terrain-referenced navigation. (c) Pendulum state estimation. (d) Battery state estimation.}
\label{application}
\end{figure}
\subsection{3D target tracking}
This example is adapted from Example 3.3.1 in \cite{exp1}. As shown in Fig. \ref{application}(a), a target moves in 3D space, and two sensors are used to track its location. The first sensor is at the origin, and the second sensor moves in a circle. The second sensor’s location $(x_{s1}, x_{s2}, x_{s3})$ at step $k=1,2,\ldots,30$ is represented by
\begin{equation}
    \begin{cases}
        x_{s1,k} = 20+20 \cos{\big(\frac{(k-1)\pi}{15}\big)}, \\
        x_{s2,k} = 20+20 \sin{\big(\frac{(k-1)\pi}{15}\big)}, \\
        x_{s3,k} = 0.
    \end{cases} 
\end{equation}
The states are the target’s positions $(x_1,x_2,x_3)$ and speeds $(v_1,v_2,v_3)$ along the three axes. The measurements $z_1,z_2$ are the distances between the target and the two sensors. The inputs are the target’s accelerations along the three axes, which are assumed to be zero in this example. When the time interval is set to $\Delta t=1$s, the system’s state transition and measurement functions can be written as (\ref{state1}) and (\ref{measure1}). Note that the noises are omitted in the equations.
\begin{equation}\label{state1}
    \begin{bmatrix}
    x_{1,k} \\
    x_{2,k} \\
    x_{3,k} \\
    v_{1,k} \\
    v_{2,k} \\
    v_{3,k}
    \end{bmatrix} 
    =
    \begin{bmatrix}
    x_{1,k-1} + v_{1,k-1}\\
    x_{2,k-1} + v_{2,k-1}\\
    x_{3,k-1} + v_{3,k-1}\\
    v_{1,k-1} \\
    v_{2,k-1} \\
    v_{3,k-1}
    \end{bmatrix} ,
\end{equation}
\begin{equation}\label{measure1}
    \begin{bmatrix}
    z_{1,k} \\
    z_{2,k} 
    \end{bmatrix} 
    =
    \begin{bmatrix}
    \sqrt{\sum_{i=1}^3x_{i,k}^2} \\
    \sqrt{\sum_{i=1}^3(x_{i,k}-x_{si,k})^2} 
    \end{bmatrix} .
\end{equation}
The true initial states of the target are $x_{1,0}=10, x_{2,0}=-10, x_{3,0}=50, v_{1,0}=1, v_{2,0}=2, v_{3,0}=0$; the initial state covariance matrix $P_0$ is diagonal, and its diagonal elements are 100, 100, 100, 0.01, 0.01, and 0.01. The initial state estimations are generated randomly from the multivariate Gaussian distribution with covariance $P_0$ and mean equal to the true initial states. The process noise covariance matrix is also diagonal, and its diagonal values are 0, 0, 0, $10^{-6}$, $10^{-6}$, and $10^{-6}$.
\subsection{Terrain-referenced navigation}
This example is adapted from Example 3.3.2 in \cite{exp1}. As shown in Fig. \ref{application}(b), a plane flies above an uneven landscape. The states are the plane’s positions along the two axes, denoted as $x_1$ and $x_2$. The measurement $z_k$ is the terrain’s elevation at the plane’s present location, which, in practice, is calculated as the difference between the barometer and radar altimeter measurements. The plane knows the contour map in advance, so the terrain’s elevation can tell the aircraft its possible locations. However, this task is difficult because all the points on a contour line have the same elevation, and system knowledge must be combined for accurate state estimations. The system inputs are the plane’s speeds along the x and y axes, which are 0.5 km/s and 0, respectively. When the time interval is set to $\Delta t=1$s, the system’s state transition and measurement functions can be written as (\ref{plane}) and (\ref{measure2}). Note that all the units for distance in this example are kilometers, and the noises are omitted in the equations.
\begin{equation}\label{plane}
    \begin{bmatrix}
    x_{1,k} \\
    x_{2,k} 
    \end{bmatrix} 
    =
    \begin{bmatrix}
    x_{1,k-1} + 0.5\\
    x_{2,k-1}
    \end{bmatrix},
\end{equation}
\begin{equation}\label{measure2}
z_k=\sin{\sqrt{\big( \frac{x_{1,k}}{40}\big)^2+\big( \frac{x_{2,k}}{40}\big)^2}}.
\end{equation}
The true initial states of the target are $x_{1,0}=10$ km, $x_{2,0}=10$ km; the initial state covariance matrix $P_0$ is diagonal, and its diagonal elements are both 1 km$^2$. The initial state estimations are generated randomly from the multivariate Gaussian distribution with covariance $P_0$ and mean equal to the true initial states. The process noise covariance matrix is also diagonal, and its diagonal values are both 0.25 m$^2$.
\subsection{Synchronous generator state estimation}
This example is adapted from a research paper \cite{generator}. The system model is precisely the same as in the paper. The states of the system are the rotor angle $\delta$, rotor speed $\Delta \omega$, the q-axis component of the voltage $e'_q$, and the d-axis component of the voltage $e'_d$. For simplicity, these states are denoted as $x_1$ to $x_4$. The measurement $z_k$ is electrical output power. The system's inputs include the mechanical input torque $T_m$, the steady-state internal voltage of the armature $E_{fd}$, and the terminal bus voltage $V_t$. For simplicity, these inputs are denoted as $u_1$ to $u_3$. The time interval is set to $\Delta t=0.1$ ms. The input profiles are the same as in the paper, which are:
\begin{equation}
    \begin{cases}
    u_{1,k}=0.8,\\
    u_{2,k}=2.11+0.0002k,\\
    u_{3,k}=1.002.\end{cases} 
\end{equation}
All the model parameters are selected to be the same as in \cite{generator}. The system's state transition and measurement functions can be written as (\ref{generatorsys}) and (\ref{measure3}). Note that the noises are omitted from the equations.
\scriptsize\begin{equation}\label{generatorsys}
    \begin{aligned}
    &\begin{bmatrix}
    x_{1,k} \\
    x_{2,k} \\
    x_{3,k} \\
    x_{4,k}
    \end{bmatrix} 
    =
    \begin{bmatrix}
    x_{1,k-1}\\
    x_{2,k-1}\\
    x_{3,k-1}\\
    x_{4,k-1}
    \end{bmatrix}+
    \begin{bmatrix}
    377x_{2,k-1}\Delta t\\
    \frac{\Delta t}{13}\big[ u_{1,k-1}-\frac{u_{3,k-1}x_{3,k-1}\sin{x_{1,k-1}}}{0.375}\big]\\
    \frac{\Delta t}{0.131}[u_{2,k-1}-x_{3,k-1}]\\
    \frac{-x_{4,k-1}\Delta t}{0.0131}
    \end{bmatrix} \\
    &+
    \begin{bmatrix}
    0\\
    \frac{\Delta t}{13}\big[0.9215u_{3,k-1}^2\sin(2x_{1,k-1})-0.05x_{2,k-1}\big]\\
    \frac{-4.4933\Delta t}{0.131}(x_{3,k-1}-u_{3,k-1}\cos{x_{1,k-1}})\\
    \frac{0.6911 u_{3,k-1}\sin{x_{1,k-1}}\Delta t}{0.0131}
    \end{bmatrix},
    \end{aligned}
\end{equation}\normalsize
\begin{equation}\label{measure3}
    z_{k}=\frac{u_{3,k-1}x_{3,k}\sin{x_{1,k}}}{0.375}-0.9215u_{3,k-1}^2\sin(2x_{1,k}).
\end{equation}
The true initial states of the system are $x_{1,0}=0.4, x_{2,0}=x_{3,0}=x_{4,0}=0$. The initial state covariance matrix $P_0$ is diagonal, and its diagonal elements are $10^{-4}$, $10^{-10}$, $10^{-4}$, and $10^{-4}$. The initial state estimations are generated randomly from the multivariate Gaussian distribution with covariance $P_0$ and mean equal to the true initial states. The process noise covariance matrix is also diagonal, and its diagonal values are $10^{-10}$, $10^{-16}$, $10^{-10}$, and $10^{-10}$.
\subsection{Pendulum state estimation}
As shown in Fig. \ref{application}(c), one end of the pendulum is fixed to the ceiling with a thin rope. The system’s states are the pendulum’s angular position  $\theta$ and angular speed $\omega$. The measurement $z_k$ is the horizontal component of tension in the rope. The system input is the external torque exerted on the pendulum, which is zero in this example. The mass of the ball is $m=1$ kg, the length of the rope is $l=1$ m, and the gravity acceleration is $g=9.8$m/s$^2$. The time interval is set to $\Delta t=0.01$s. According to Newton’s law, when the time interval is small, the system's state transition and measurement functions can be written as (\ref{pendulumsys}) and (\ref{measure4}). Note that the noises are omitted in the equations.
\begin{equation}\label{pendulumsys}
    \begin{bmatrix}
    \omega_{k} \\
    \theta_{k} 
    \end{bmatrix} 
    =
    \begin{bmatrix}
    \omega_{k-1} - \frac{g}{l}\sin(\theta_{k-1})\Delta t\\
    \theta_{k-1} + \omega_{k-1}\Delta t
    \end{bmatrix} ,
\end{equation}
\begin{equation}\label{measure4}
z_k=mg\cos{\theta_k}\sin{\theta_k}+ml\omega_k^2\sin{\theta_k}.
\end{equation}
The true initial states of the system are $\omega_0=0,\theta_0=\pi/4$. The initial state covariance matrix $P_0$ is diagonal, and its diagonal elements are both $(\pi/18)^2$. The initial state estimations are generated randomly from the multivariate Gaussian distribution with covariance $P_0$ and mean equal to the true initial states. The process noise covariance matrix is also diagonal; its diagonal values are $10^{-10}$ and 0.
\subsection{Battery state estimation}
This example is adapted from a research paper \cite{battery}. As shown in Fig. \ref{application}(d), a battery can be modeled as several circuit components, including a voltage source OCV (stands for the cell's open-circuit voltage), an internal resistor $R_1$, a transfer resistor $R_2$, and a capacitor $C_1$. A battery's two most important states are its state of charge (SOC) and state of health (SOH). The SOC is the ratio of the battery’s remaining capacity to its present maximum capacity, and the SOH is the ratio of the battery’s present maximum capacity to its initial maximum capacity. In general, SOC describes the battery's energy level, while SOH describes the battery’s aging level. The states of this system are the SOC, SOH, and capacitor voltage $U_c$. The measurement $z_k$ is the battery’s terminal voltage. The input is the charging current, which, in this example, is a three-level square wave shown below:
\begin{equation}
I_k =
\begin{cases}
-2, & 15 < k \le 75,\\
 2, & 105 < k \le 165,\\
 0, & \text{else}.
\end{cases}
\end{equation}
The battery’s initial maximum capacity is $Q_0=1\,\mathrm{Ah}=3600\,\mathrm{A\,s}$. The parameters in the equivalent circuit model are chosen to be $R_1=0.01\,\Omega, R_2=0.05\,\Omega, 1/(R_2C_1)=0.008\,\mathrm{s}^{-1}$. The OCV is a function of SOC and SOH, and it can be written as:
{\scriptsize\begin{equation}
    OCV=\sum_{i=0}^9 \big( \frac{SOH-0.8}{0.2}a_{100,10-i}+\frac{1-SOH}{0.2}a_{80,10-i}\big)SOC^i,
\end{equation}}\normalsize	
where $a_{100}= [1390.38, -6961.31, 14760.31, -17230.92,$ $12055.71, -5162.75, 1330.60, -196.37, 15.60, 2.96]$, $a_{80}=$ $[813.94, -4229.96, 9345.49, -11415.38, 8396.15, -3801.07,$ $1043.09, -165.29, 14.28, 2.96]$. The time interval is set to $\Delta t=1$s, and the total simulation time is 180 seconds. The system's state transition and measurement functions can be written as \eqref{batterysys} and \eqref{measure5}. Note that the noises are omitted in the equations.
\begin{equation}\label{batterysys}
    \begin{bmatrix}
    SOC_{k} \\
    U_{c,k} \\
    SOH_{k}
    \end{bmatrix}
    =
    \begin{bmatrix}
    SOC_{k-1}+\frac{I_{k-1}\Delta t}{Q_0 SOH_{k-1}}\\
    \alpha U_{c,k-1}+(1-\alpha)R_2I_{k-1} \\
    SOH_{k-1}
    \end{bmatrix}
\end{equation}
\begin{equation}\label{measure5}
z_k=OCV(SOC_{k},SOH_k)+U_{c,k}+R_1I_k
\end{equation}
where $\alpha=\exp{\big(\frac{-\Delta t}{R_2C_1}\big)}$.
The true initial states of the system are $SOC_0=60\%, U_{c,0}=0, SOH_0=90\%$. The initial state covariance matrix $P_0$ is diagonal; its diagonal elements are 0.04, $10^{-10}$, and 0.01. The initial state estimations are 80\%, 0, and 100\%, respectively. The input current is noisy, and its standard deviation is 1 mA. This input noise is the only source of the process noises. Since the input current also appears in the measurement function, the actual measurement noise is the combination of the measurement noise of the voltmeter and the noise caused by the term $R_1I_k$ in the measurement function.
\end{document}